\documentclass[useAMS,usenatbib,twocolumn,fleqn]{mnras}
\usepackage[T1]{fontenc}
\usepackage{ae,aecompl}
\usepackage{natbib, graphics, epsfig}
\usepackage{stmaryrd}
\usepackage{times}
\usepackage{bigints}

\usepackage{times}
\usepackage{amsmath}
\usepackage{amssymb}
\usepackage{textcomp}
\usepackage{verbatim}
\usepackage{enumitem}

\newcommand{\msun}{{\,\rm M_\odot}}

\newcommand{\kms}{\,{\rm km}\,{\rm s}^{-1}}
\newcommand{\cm}{\,{\rm cm}}

\newcommand*\diff{\mathop{}\!\mathrm{d}}
\newcommand{\Gyr}{\,{\rm Gyr}}
\newcommand{\K}{\,{\rm K}}
\newcommand{\Myr}{\,{\rm Myr}}
\newcommand{\pc}{\,{\rm pc}}
\newcommand{\kpc}{\,{\rm kpc}}
\newcommand{\Mpc}{\,{\rm Mpc}}

\newcommand{\gcm}{\, {\rm g}\,{\rm cm}^{-3}}
\newcommand{\mum}{\,\mu{\rm m}}

\let\baraccent=\= 
\renewcommand{\=}[1]{\stackrel{#1}{=}} 

\usepackage{color}

\def\aap{A\&A}
\def\apj{ApJ}

\def\apjl{ApJ}
\def\mnras{MNRAS}
\def\araa{ARA\&A}

\def\nat{Nature}

\title[Dust in cluster environments and its impact on gas cooling]
      {Dust in and around galaxies: dust in cluster environments and its impact on gas cooling}

      \author[M. Vogelsberger et al.] {\parbox{18.5cm}{
	  Mark Vogelsberger$^1$\thanks{email:mvogelsb@mit.edu},
          Ryan McKinnon$^1$,
	  Stephanie O'Neil$^1$,
	  Federico Marinacci$^{1,2}$,\\
	  Paul Torrey$^{1,3}$ and Rahul Kannan$^{1,2}$\thanks{Einstein Fellow} 
        }\vspace{0.3cm}\\
        $^1$ Department of Physics, Kavli Institute for Astrophysics and Space Research, Massachusetts Institute of Technology, Cambridge, MA 02139, USA\\
        $^2$ Harvard-Smithsonian Center for Astrophysics, 60 Garden Street, Cambridge, MA, 02138, USA\\
        $^3$ Department of Astronomy, University of Florida, 211 Bryant Space Sciences Center, Gainesville, FL 32611 USA
}

\date{Accepted XXX. Received YYY; in original form ZZZ}

\pubyear{2018}

\begin{document}
\label{firstpage}
\pagerange{\pageref{firstpage}--\pageref{lastpage}}
\maketitle





\begin{abstract}
Simulating the dust content of galaxies and their surrounding gas is
challenging due to the wide range of physical processes affecting the dust
evolution. Here we present cosmological hydrodynamical simulations of a cluster
of galaxies, $M_\text{200,crit}=6 \times 10^{14}\msun$, including a novel dust
model for the moving mesh code {\sc Arepo}. This model includes dust
production, growth, supernova-shock-driven destruction, ion-collision-driven
thermal sputtering, and high temperature dust cooling through far infrared
re-radiation of collisionally deposited electron energies. Adopting a rather low thermal sputtering rate, we find, consistent with
observations, a present-day overall dust-to-gas ratio of $\sim
2\times 10^{-5}$, a total dust mass of $\sim 2\times 10^9\msun$, and a dust
mass fraction of $\sim 3\times 10^{-6}$. The typical thermal sputtering
timescales within $\sim 100\kpc$ are around $\sim 10\Myr$, and increase towards
the outer parts of the cluster to $\sim 10^3\Myr$ at a cluster-centric distance
of $1\Mpc$. The condensation of gas phase metals into dust grains reduces high
temperature metal-line cooling, but also leads to additional dust infrared
cooling. The additional infrared cooling changes the overall cooling rate in
the outer parts of the cluster, beyond $\sim 1\Mpc$, by factors of a few.  This
results in noticeable changes of the entropy, temperature, and density profiles
of cluster gas once dust formation is included. The emitted dust infrared
emission due to dust cooling is consistent with observational constraints.
\end{abstract}

\section{Introduction}
\label{sec:Section1}

Dust is an integral component of the galactic ecosystem and is crucial for a
plethora of physical processes in the interstellar medium (ISM). Within the
ISM, dust undergoes different surface reactions and acts as a catalyst for the
formation of molecules~\citep[][]{Hollenbach1971, Mathis1990, Li2001,
Draine2003}. Gas phase metals condense onto dust grains, which leads to the
depletion of the gas phase metal budget in the ISM~\citep[][]{Calzetti1994,
Calzetti2000, Netzer2007, Spoon2007}.  The actual dust mass of a galaxy depends
on its properties and also its redshift. For the Milky Way about $\sim 50\%$ of
the metal mass is locked into the dust component. This amounts to $1\%$ of the
total mass budget in the ISM.  Furthermore, dust grains absorb stellar
radiation in the ultraviolet and re-emit this radiation in the infrared
(IR)~\citep[][]{Spitzer1978, Draine1984, Mathis1990, Tielens2005}. The presence
of cosmic dust is inferred through its IR emission or reddening of stellar
light.  An observational challenge is its detection around galaxies and
especially in galaxy clusters and in the intracluster medium
(ICM)~\citep[e.g.,][]{PlanckCollaboration2016, Erler2018, Melin2018}.
Information about the dust content in the ICM would lead to insights into dust
production, destruction, dust cooling mechanisms, and gas and dust stripping
from galaxies.  More generally, there is a strong interest in quantifying the
amount and distribution of dust around galaxies outside of the
ISM~\citep[e.g.,][]{Menard2010}.

\begin{table*}
\begin{center}
\begin{tabular}{llllll}
\hline
Model Name                          & Dust Thermal Sputtering                                            & Dust IR Cooling                                      & Dust Grain Size                       & Comment\\
\hline
\hline
{\small\sc no-dust}                 & --                                                                 & --                                                   & --                                    & no dust included  \\
\hline
{\small\sc fiducial}                & fiducial                                                           & fiducial                                             & fiducial (${a=0.1\mum}$)                             & \protect\cite{McKinnon2016} \\&&&& w/ reduced thermal sputtering \\ &&&& + dust growth only in star-forming gas \\ &&&& + dust IR cooling\\
\hline
{\small\sc slow-sputter}            & ten times slower ($10 \times \tau_\text{sputter}^\text{ref}$)     & fiducial                                             & fiducial                              & strongly reduced thermal sputtering\\
\hline
{\small\sc large-grains}            & fiducial                                                           & fiducial                                             & five times larger ($5 \times a$)       & larger dust grains\\
\hline
{\small\sc more-cooling}            & fiducial                                                           & five times more ($5 \times \Lambda_\text{dust}$)      & fiducial                              & enhanced dust IR cooling \\
\hline
\end{tabular}
\end{center}
\caption{{\bf Summary of dust model variations explored in this work.} We vary
three model ingredients that are regulating the amount of dust in
the simulated cluster and its impact on the gas cooling: the thermal sputtering
timescale, the strength of the IR cooling rate, and the dust grain size. The
{\small\sc slow-sputter} model is an extreme case with a very low sputtering
rate to demonstrate an absolute upper limit for the amount of dust in the cluster and its ICM.
The other model variations are more realistic given the uncertainties for the
various processes.}
\label{table:dust_models}
\end{table*}

The overall cluster dust-to-gas mass ratio, $D={M_\text{dust}/M_\text{gas}}$,
is not well constrained.  \cite{Chelouche2007} found that the dust-to-gas ratio in clusters should be less than $5\%$ of the
local ISM value of $\sim 10^{-2}$ based on extinction studies. \cite{Giard2008} found ${D = 5\times
10^{-4}}$ if all their detected IR luminosity towards galaxy clusters is
produced by thermal emission from ICM dust.  This is close to ${D=3\times
10^{-4}}$ as reported in~\cite{McGee2010}.  By  modelling IR properties of the
galactic population of the SDSS-maxBCG clusters \cite{Roncarelli2010} found an
upper limit of $D\lesssim 5\times 10^{-5}$. The \cite{PlanckCollaboration2016}
used cluster stacking of IR spectral energy distributions to infer ${D = (1.93
\pm 0.92) \times 10^{-4}}$ for their full cluster sample (${\langle z \rangle =
0.26 \pm 0.17}$).  For the low- ($z <0.25$) and high-redshift ($z > 0.25$)
sub-samples, they found ${D=(0.79 \pm 0.50) \times 10^{-4}}$ and ${D=(3.7 \pm
1.5) \times 10^{-4}}$, respectively. They also identified a trend with halo
mass, where ${D=(0.51 \pm 0.37) \times 10^{-4}}$ (${M_\text{500} < 5.5\times
10^{14}\msun}$) and ${D=(4.6 \pm 1.5) \times 10^{-4}}$ (${M_\text{500} >
5.5\times 10^{14}\msun}$). \cite{Kitayama2009} searched for IR emission within
the Coma cluster and found an upper limits of $D=10^{-5}$ within the central
$100\kpc$ by masking out IR point sources. \cite{Gutierrez2014} found a dust mass ratio within their cluster
sample of $9.5 \times 10^{-6}$ and a dust-to-gas ratio about three orders of
magnitude lower than the value found in the Milky Way. Despite the measurement
uncertainties and differences between all these studies, it is evident that
there is most likely only a small amount of dust present in the ICM. It is
expected that the hot ICM environment very efficiently destroys dust and
therefore causes an overall low abundance of dust in the ICM.

So far, only a limited number of theoretical studies have tried to quantify the
dust content within groups and clusters, which is mainly due to the lack of
detailed dust models.  Some simple dust models have predicted that the mass
fraction of dust in clusters can reach $1-3\%$ of the galactic
value~\citep{Polikarpova2017}.  \cite{Masaki2012} predicted dust mass fractions
in groups to be of the order of $10^{-5}$. More recently, \cite{Gjergo2018}
presented the most detailed dust calculation of a galaxy cluster so far using a
combination of a dust model coupled to a galaxy formation model. 
They studied four clusters adopting a two size grain approximation, and predicted
a dust content that is largely consistent with the measurements of the
\cite{PlanckCollaboration2016}.  Interestingly, they had to increase the
thermal sputtering timescale of their fiducial model by a factor of $5$ to match the
observed dust content of clusters.

Once dust grains are produced and exist, they also act as a heating source or
coolant depending on the physical state of the surrounding gas and radiative
environment.  Heating operates via the photoelectric effect if the stellar
radiation field is strong enough, and high temperature dust cooling occurs through IR re-radiation of collisionally
deposited energy on grains by impinging free
electrons~\citep[][]{Ostriker1973}. Unfortunately, only little is known about the importance
of this cooling channel within galaxy clusters. \cite{Montier2004} predicted
that dust cooling is important in the ICM for gas temperatures ${T_\text{gas} = 10^6 -
10^8\K}$ and if $D > 2\times 10^{-5}$. \cite{DaSilva2009} found a $25\%$
normalisation change for the $L_\text{X}-M$ relation and a $10\%$ change for
the $Y-M$ and $S-M$ cluster scaling relations in the presence of dust.
Similarly, \cite{Pointecouteau2009} found changes in the $L_\text{X}-M$
relation by as much as $10\%$ for clusters with temperatures around $1\,{\rm
keV}$ for models that include dust cooling.  However, those results are based
on simplified dust models and it is currently unclear whether they capture the
correct physical behaviour. A more detailed inclusion of dust physics has only
recently been achieved in galaxy formation
simulations~\citep[e.g.,][]{Bekki2015, McKinnon2016, Zhukovska2016, Aoyama2017, Popping2017, McKinnon2017,Aoyama2018, Gjergo2018,McKinnon2018}. However, none of these models included
the effect of dust cooling so far.

In this paper we present cosmological simulations of dust within galaxy
clusters and the ICM using a self-consistent model for dust physics including
dust production, growth, destruction, thermal sputtering and IR cooling. The
scope of our work is similar to that of~\cite{Gjergo2018}, who recently studied
the global dust content of galaxy clusters through simulations. Here we focus
also on the spatial distribution of dust within the cluster, and we also study its impact on
the thermodynamic properties of the ICM gas by including dust IR cooling.
Our paper is structured as follows. In Section~\ref{sec:Section2} we present our model and
simulation details. In Section~\ref{sec:Section3} we then discuss the global dust content of the
cluster and compare our predictions with observational constraints. In the
following Section~\ref{sec:Section4} we study the distribution of dust in the cluster gas. In
Section~\ref{sec:Section5} we then explore how and whether dust can affect the thermodynamics of
the gas in clusters. We give our conclusions in Section~\ref{sec:Section6}.

\section{Methods}
\label{sec:Section2}

We simulate a galaxy cluster, $M_\text{200,crit}=6 \times 10^{14}\msun$,
based on zoom-in initial conditions using the moving-mesh {\sc Arepo} code~\citep[][]{Springel2010} combined with the IllustrisTNG galaxy formation model~\citep[][]{Weinberger2017, Pillepich2018},
which is an update of the original Illustris model~\citep[][]{Vogelsberger2013, Torrey2014, Vogelsberger2014a, Vogelsberger2014b}. This model is complemented by
a novel dust model~\citep[][]{McKinnon2016, McKinnon2017} with additional far
IR dust cooling. The cosmological parameters of the simulation are: $\Omega_m =
0.3089$, $\Omega_\Lambda = 0.6911$, $\Omega_b = 0.0486$, $\sigma_8 = 0.8159$,
$n_s = 0.97$, and ${H_0 = 67.74\, \rm{km \, s^{-1} Mpc^{-1}}}$. The high
resolution dark matter and gas masses of our zoom-in simulation are  ${1.2
\times 10^7\msun}$ and ${1.9 \times 10^6\msun}$ respectively with a dark matter softening
length of $1.4\kpc$ and an adaptive gas cell softening.

The dust is modelled and followed using a fluid passive scalar that evolves
according to characteristic timescales for different physical processes such
that the dust mass, $M_\text{dust}$, within a gas cell evolves
as~\citep[][]{McKinnon2016, McKinnon2017}:
\begin{equation}
\frac{\diff M_\text{dust}}{\diff t} \!=\! \left(\! 1 \!-\! \frac{M_\text{dust}}{M_\text{metal}} \right) \! \left( \frac{M_\text{dust}}{\tau_\text{growth}} \right) - \frac{M_\text{dust}}{\tau_\text{SNII shocks}} \! - \! \frac{M_\text{dust}}{\tau_\text{sputter}},\nonumber
\end{equation}
where $M_\text{metal}$ is the mass in gas phase metals in the cell. The dust
mass evolution is determined by three timescales associated with different
physical processes. The growth timescale, $\tau_\text{growth}$, the dust
destruction timescale due to type II supernovae (SNeII), $\tau_\text{SNII
shocks}$, and the dust destruction timescale due to thermal sputtering,
$\tau_\text{sputter}$. The first factor in the parentheses 
depends on the local dust-to-metal ratio and slows the
accretion rate down as gas-phase metals are condensed into dust.

\begin{figure*}
\centering
\includegraphics[width=1\textwidth]{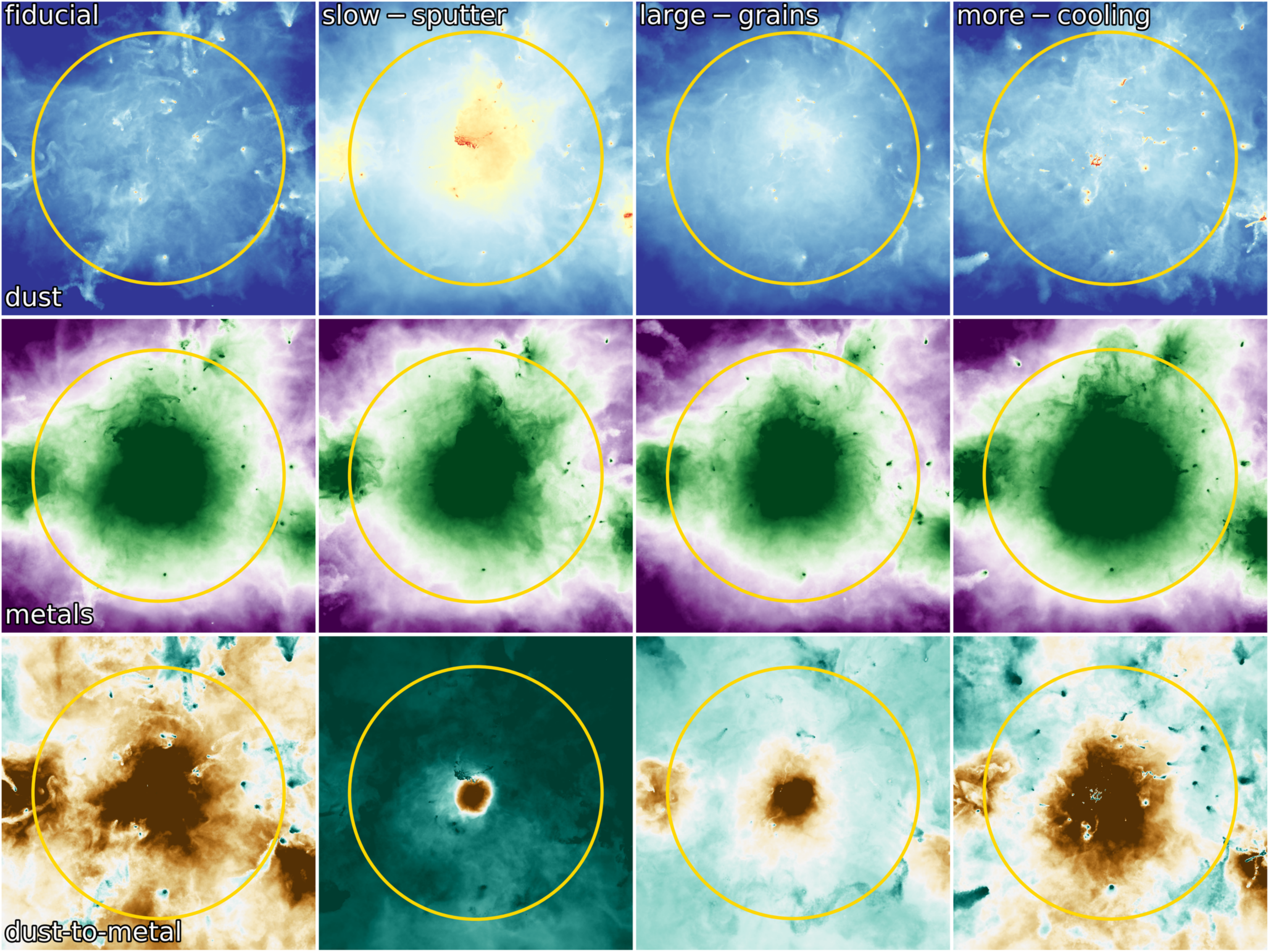}
\includegraphics[width=0.235\textwidth]{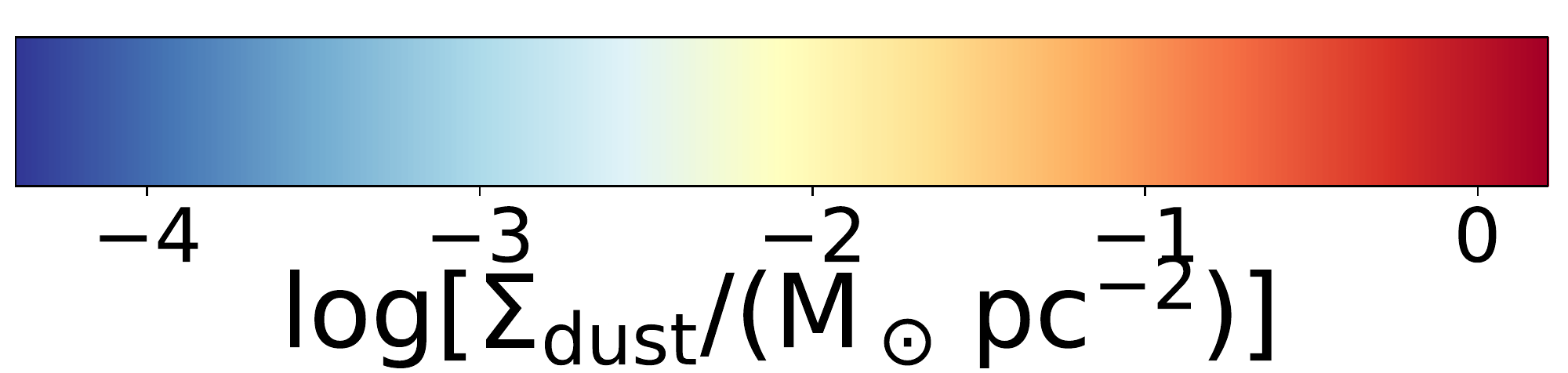}
\hspace{2cm}\includegraphics[width=0.235\textwidth]{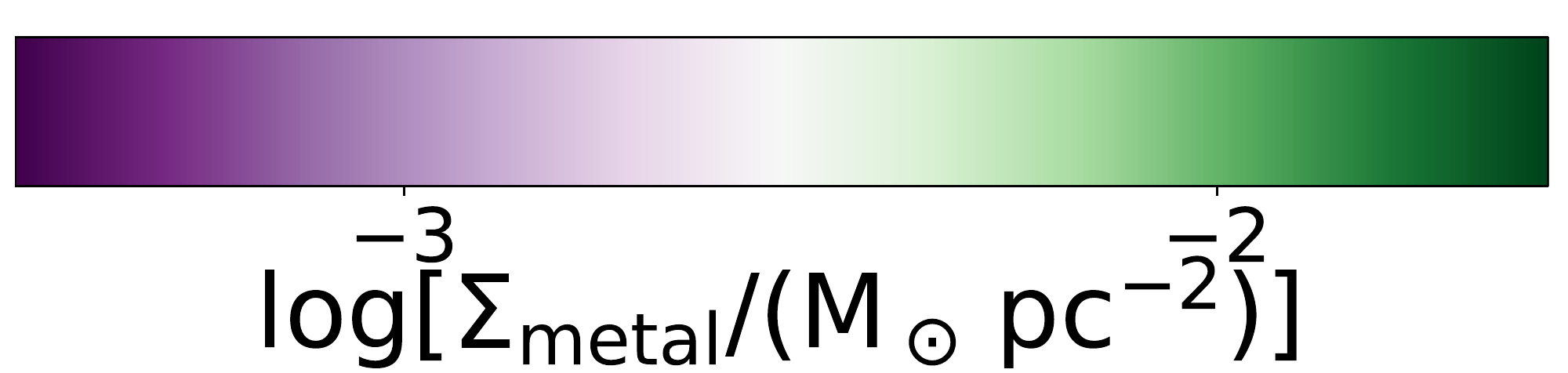}
\hspace{2cm}\includegraphics[width=0.235\textwidth]{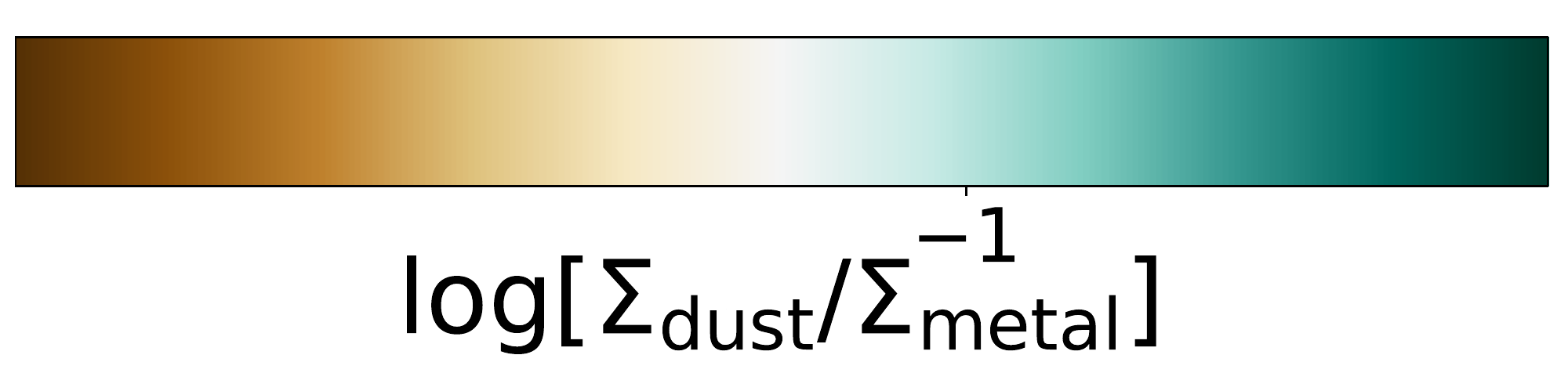}
\caption{{\bf Dust maps for different dust models at $\mathbf{z=0}$.} {\it Top
row:} Dust surface density maps. {\it Middle row:} Metal surface density maps.
{\it Bottom row:} Dust-to-metal ratio maps. The circles show $r_{\rm 200,
crit}$. The model with slow thermal sputtering, {\small\sc slow-sputter}, leads
to an increased amount of dust in the cluster, where dust closely follows the
distribution of metals in the ICM due to the lack of efficient dust destruction
for that model. This also leads to a rather constant dust-to-metal ratio
throughout the ICM. The {\small\sc large-grains} and {\small\sc more-cooling}
models also lead to slightly larger amounts of dust in the cluster compared to the {\small\sc
fiducial} model. For the {\small\sc fiducial} model, the dust-to-metal ratio
can fluctuate by more than one order of magnitude demonstrating that metals
are generally an unreliable tracer of dust. This is mainly because metals do
not experience thermal sputtering in the ICM, and metals also do not experience
growth processes in the ISM. }
\label{fig:dust_maps_1}
\end{figure*}

The production of dust is coupled to the stellar evolution implementation of
our galaxy formation model, where we assume that some of the metals condense
into dust grains based on specific dust condensation efficiency
parameters~\citep[see][for details]{McKinnon2016, McKinnon2017}.  We adopt different parametrisations for the amount of dust produced from asymptotic giant
branch (AGB) stars and from SNe.  Additionally, we make a distinction between
AGB stars with $\rm{C/O} > 1$ in their stellar envelope, which are expected to
produce carbonaceous solids (e.g., graphite or amorphous carbon), and those
with ${\rm C/O} < 1$, which are thought to form primarily silicate dust. Once
the dust has been produced during stellar evolution, the time evolution of its
mass budget follows the mass equation above assuming that dust behaves as a
passive scalar within the underlying fluid.  In the following, we briefly
describe the different physical processes that are relevant for the dust
evolution once the grains have been produced. We specify the relevant
timescale for each process.\\

\noindent {\it Dust growth:} Dust grains in the ISM gain mass when gas atoms
collide with them and stick onto their surfaces~\citep[][]{Draine1990}. The dust
growth timescale due to gas condensation onto existing dust grains is given by:
\begin{equation}
\tau_\text{growth} = \tau_\text{growth}^\text{ref} \, \frac{a}{0.1\mum} \left( \frac{\rho_\text{gas,growth}^\text{ref}}{\rho_\text{gas}} \right) \left( \frac{T_\text{gas,growth}^\text{ref}}{T_\text{gas}} \right)^{1/2}, \nonumber
\end{equation}
where $a$ is the dust grain size, $T_\text{gas}$ is the gas temperature, and $\rho_\text{gas}$ is the
gas mass density~\citep[][]{Dwek1998, Hirashita2000}. This growth timescale is
shortest in dense gas where dust-gas collisions are more frequent. We slightly
modify the original growth prescriptions of \cite{McKinnon2016} such that dust
can only grow in star-forming gas. $\tau_\text{growth}^\text{ref}$ is a
normalisation constant.\\

\noindent{\it Dust destruction due to SNII shocks:}  Blast waves from SNeII
produce harsh environments for dust that shrink dust grains and cause them to
lose mass.  The dust SNII shock destruction timescale for a gas cell of mass $M_\text{gas}$ is given by:
\begin{equation}
\tau_\text{SNII shocks} = \frac{M_\text{gas}}{\epsilon \, \gamma \, M_\text{s}(100)}, \nonumber
\end{equation}
where $\epsilon$ is the dust destruction efficiency, $\gamma$ is the local SNII
rate, and $M_\text{s}(100)$ is the mass of gas shocked to at least
$100\kms$~\citep[][]{Dwek1980, McKee1989} calculated using the Sedov-Taylor
solution of a homogeneous environment.\\

\noindent {\it Dust destruction due to thermal sputtering:} At high
temperatures, gas ions have large thermal velocities and can collisionally
erode dust grains.  For dust in the ICM, this process of thermal sputtering
plays a crucial role since it is the main destruction mechanism in this hot
environment \citep[e.g.,][]{Draine1979b}. The thermal sputtering timescale for
dust in hot gas can be approximated by~\citep[][]{Tsai1995}:
\begin{equation}
\tau_\text{sputter} \! = \! \tau_\text{sputter}^\text{ref} \,\!\!  \left( \frac{a/0.1\mum}{\rho_\text{gas}/10^{-27}\gcm} \right) \! \left[ \!\! \left( \frac{2\times 10^6\,\text{K}}{T_\text{gas}} \right)^{2.5} \!\!\!\!\!\!\!\! + 1  \right], \nonumber
\end{equation}
where $\tau_\text{sputter}^\text{ref}$ is a normalisation constant. This
sputtering timescale parametrisation has also been used in the recent dust model
of~\cite{Gjergo2018}.\\

\noindent{\it High temperature dust cooling:} Hot gas plasma electrons that collide
with dust grains can lose energy and cool, while the dust grains heat up and subsequently
radiate this energy away in the IR. The original model of~\cite{McKinnon2016,
McKinnon2017} did not include any heating or cooling effects due to dust. Here
we include dust cooling due to IR radiation in high temperature gas
environments such as the ICM. 

The electron collisional heating rate for a single grain of radius $a$ in gas of
temperature $T_\text{gas}$ and electron density $n_\text{e}$ is given by
$H(a, T_\text{gas}, n_\text{e})= n_\text{e} \, \widetilde{H}(a, T_\text{gas})$ with~\citep[][]{Dwek1981,Dwek1987}:
\begin{equation}
\frac{\widetilde H(a, T_\text{gas})}{\text{erg} \, \text{s}^{-1} \, \text{cm}^3}  \! = \! \left\{
\begin{array}{ll}
\!\!\!\! 5.38 \!\! \times \!\! 10^{-18} \Big( \frac{a}{\mu\text{m}} \Big)^{2} \! \Big( \frac{T_\text{gas}}{\text{K}} \Big)^{1.5},      x \geq 4.5         ,\\[0.2cm]
\!\!\!\! 3.37 \!\! \times \!\! 10^{-13} \Big( \frac{a}{\mu\text{m}} \Big)^{2.41} \! \Big( \frac{T_\text{gas}}{\text{K}} \Big)^{0.88},  1.5 \leq x < 4.5   ,\\[0.2cm]
\!\!\!\! 6.48 \!\! \times \!\! 10^{-6}  \Big( \frac{a}{\mu\text{m}} \Big)^{3},                                               x < 1.5         , 
\end{array}\nonumber
\right.
\end{equation}
where ${x = 2.71 \times 10^{8} (a/\mu\text{m})^{2/3}
(T_\text{gas}/\text{K})^{-1}}$. We note that we neglect heating through proton collisions and
radiation since those are subdominant in the ICM~\citep[][]{Montier2004}. The
volumetric gas cooling rate, $\Lambda_\text{dust}$ with ${[\Lambda_\text{dust}]={\rm erg}\,{\rm s}^{-1}\,{\rm cm}^{-3}}$, due to dust heating is then
given by
\begin{align}
\Lambda_\text{dust}(T_\text{gas}, D, n_\text{e}) &= n_\text{dust} \, n_\text{e} \widetilde{H} (a, T_\text{gas}) \nonumber
\end{align}
where 
\begin{equation}
{n_\text{dust} =D \, \frac{\rho_{\rm gas}}{m_\text{dust}}} = \left (\frac{D \, m_\text{p}}{X \,m_\text{dust}} \right )\,n_\text{H}  \nonumber
\end{equation}
is the number density of dust grains for a given dust-to-gas ratio $D$, $m_\text{dust}$ is the grain mass, and $X$ is the hydrogen mass fraction. Here, ${m_\text{dust} = 4 \pi /3 \rho_\text{grain} a^3}$ is calculated using the internal grain density ${\rho_\text{grain} = 3 \, \text{g} \, \text{cm}^{-3}}$.
We can turn this volumetric cooling rate into a cooling function 
\begin{align}
\frac{\Lambda_\text{dust}(T_\text{gas}, D, n_\text{e})}{n_\text{H}^2} &= \left(\frac{n_\text{dust}}{n_\text{H}}\right)  \left(\frac{n_\text{e}}{n_\text{H}}\right)  \widetilde{H}(a,T_\text{gas}) \nonumber \\  \nonumber
&= \left (\frac{D \, m_\text{p}}{X \, m_\text{dust}} \right) \left(\frac{n_\text{e}}{n_\text{H}}\right)  \widetilde{H} (a, T_\text{gas}) \\ \nonumber
&= \left( \frac{D\, m_\text{p} }{X\, 4\pi/3 \, a^3 \, \rho_\text{grain}} \right) \left(\frac{n_\text{e}}{n_\text{H}} \right)  \widetilde H(a, T_\text{gas}).
\end{align}
with ${[\Lambda_\text{dust}/n_\text{H}^2] = {\rm erg}\,{\rm s}^{-1}\,{\rm cm}^3}$. 
We note that for a hot plasma we have $n_\text{e}/n_\text{H}\cong1+Y/(2X)$, where $Y$ is the helium mass fraction. The
dust IR cooling rate therefore scales linearly with the dust-to-gas ratio, $D$,
and depends on the grain size such that larger grains typically lead to
less cooling. We implement this cooling function in {\sc Arepo} in addition
to the primordial and metal line cooling.\\

\begin{figure*}
\centering
\begin{tabular}{ll}
\includegraphics[width=0.49\textwidth]{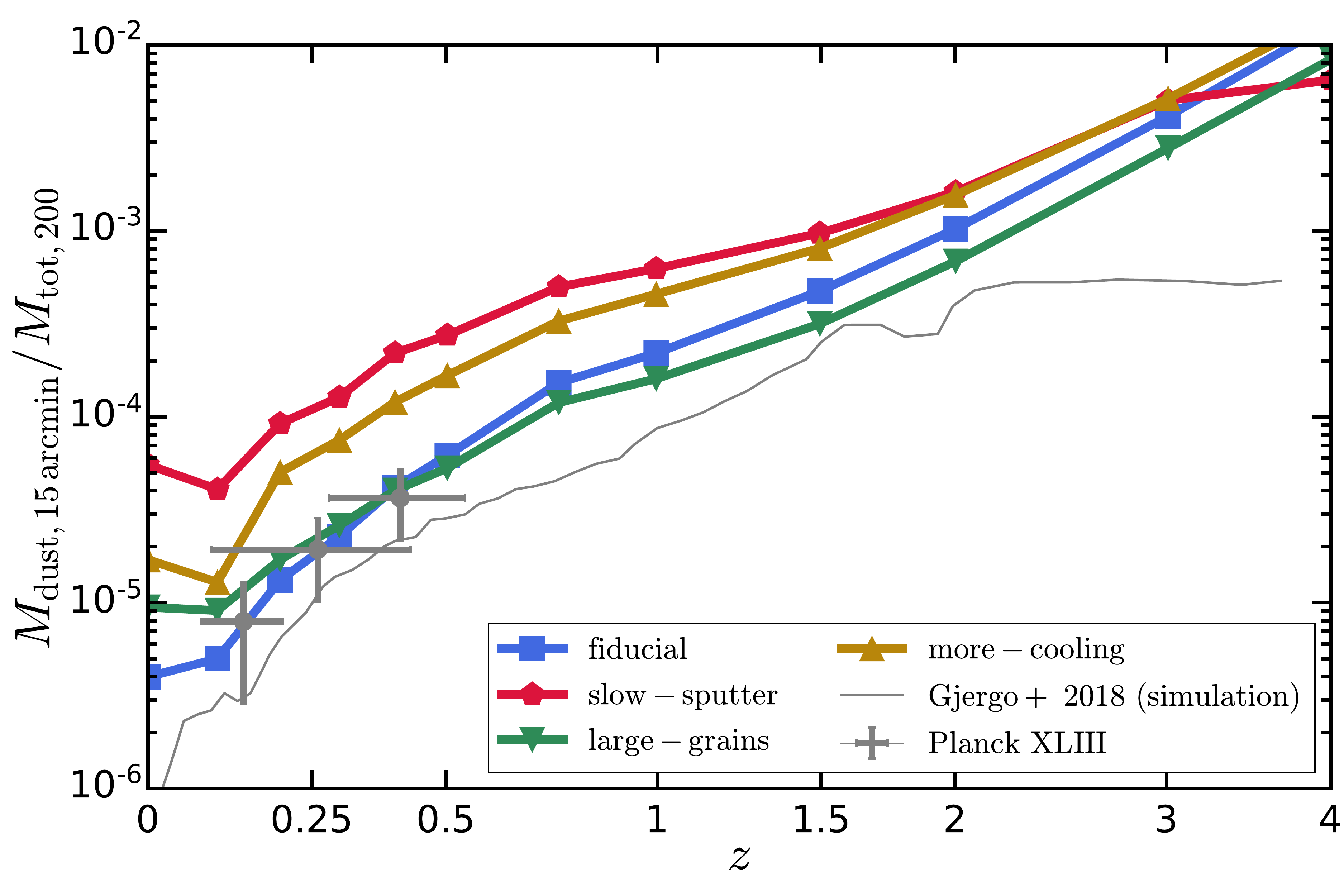}
\includegraphics[width=0.49\textwidth]{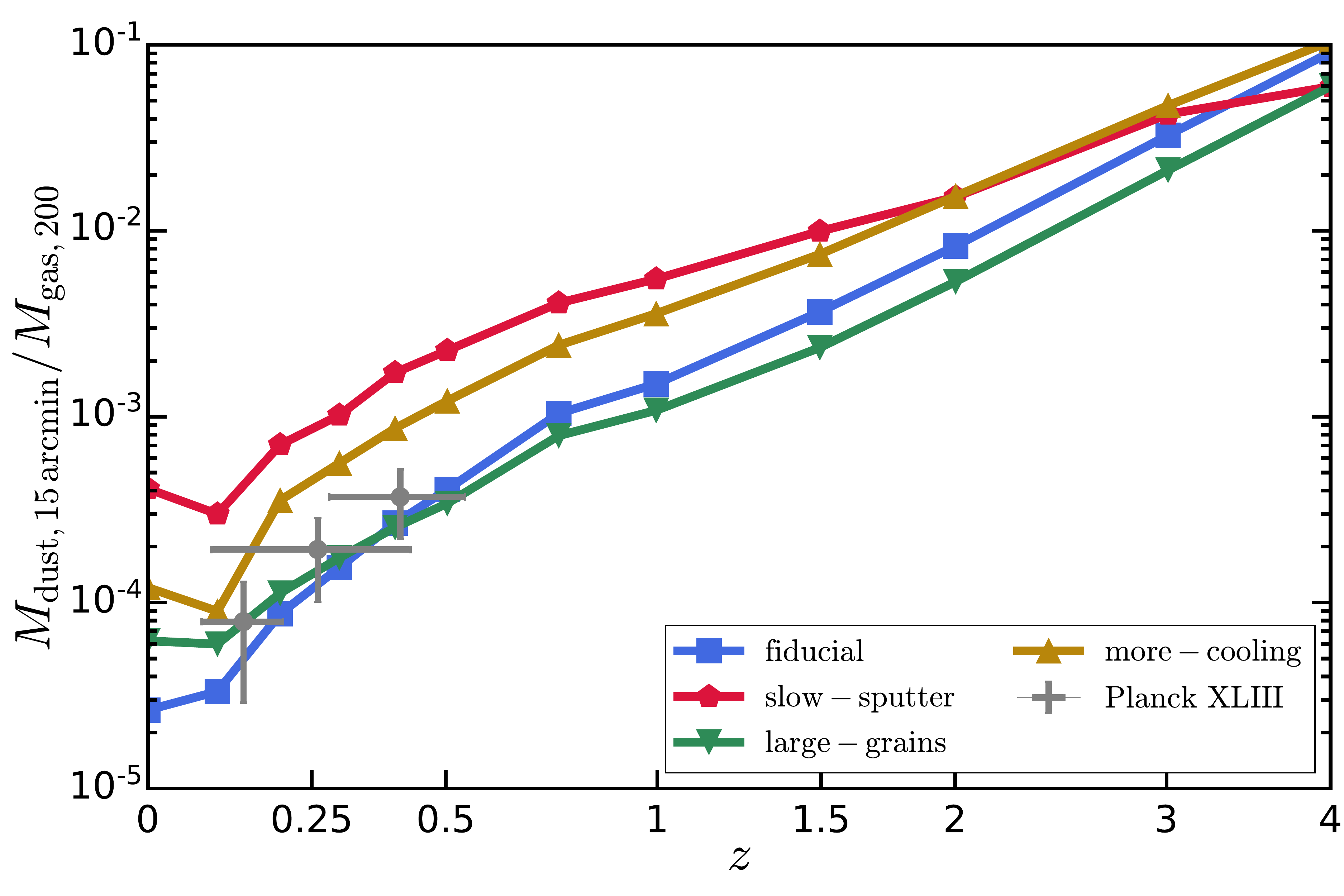}
\end{tabular}
\caption{{\bf Redshift evolution of the total dust mass contained in the
cluster.} {\it Left panel:} Total dust mass within $15\,{\rm arcmins}$ at each
redshift. At lower redshift, once $15\,{\rm arcmins}$ is smaller than $r_{\rm
200, crit}$, we replace this radial cut with $r_{\rm 200, crit}$. We divide the
dust mass at all redshifts by the total cluster mass within $r_{\rm 200,
crit}$.  {\it Right panel:} Same as the left panel, but normalising the dust
mass to the gas mass within $r_{\rm 200, crit}$. Towards lower redshifts we
also include observational data from the
\protect\cite{PlanckCollaboration2016}.  Our {\small\sc fiducial} model is
consistent with those observations.}
\label{fig:dust_time_evolution_1}
\end{figure*}

\noindent {\it Fiducial dust model, model uncertainties and model variations:}
Our dust model includes some free and adjustable parameters, which have to be
set to certain values. Our {\small\sc fiducial} dust model parameters
are~\citep[][]{McKinnon2016,McKinnon2017}: ${a=0.1\mum}$ (typical ICM grains
have ${0.03\mum < a < 0.2 \mum}$~\citep[][]{Ferrara1991}),
${\tau_\text{growth}^\text{ref}=0.2\Gyr}$, ${T_\text{gas,growth}^\text{ref}=20\K}$,
${\rho_\text{gas, growth}^\text{ref}=2.3\times10^{-24}\gcm}$,
$\tau_\text{sputter}^\text{ref}=0.57\Gyr$, and $\epsilon=0.3$.  We note that
this model employs a sputtering timescale ten times larger than the fiducial
model originally discussed in~\cite{McKinnon2017}. We have found that this
increase is required to match the observed dust abundance in the cluster.
Higher sputtering rates result in significantly too low dust masses, which are
in tension with current observational estimates. We note that \cite{Gjergo2018}
came to a similar conclusion as we will discuss in more detail below. 

The parametrisations of the different physical processes have some
uncertainties. For example, we assume constant
sticking efficiencies for our dust growth timescale, which can lead to growth rate
variations~\citep{Zhukovska2016}. The thermal sputtering timescale is also
modelled using a fit~\citep[][]{Tsai1995} to ab initio calculations, which
causes inaccuracies of around $\sim1.5 \, \text{dex}$ depending on grain
composition and temperature \citep[e.g.,][]{Barlow1978, Draine1979b}. This
inaccuracy is potentially also the reason the fiducial thermal sputtering
fit employed in~\cite{McKinnon2017} leads to too much dust destruction, and we must employ a ten times higher normalisation for the sputtering
timescale in this work. Our shock destruction implementation depends on the
destruction efficiency $\epsilon$, which is expected to be in the range from $0.1$ to $0.5$
\citep[][]{McKee1989} and is therefore also not certain.  The exact IR
dust cooling rates also depend on the detailed composition of the
dust~\citep[e.g.,][]{DaSilva2009}, which we do not track in detail in our
simulations. This introduces additional uncertainties in the cooling rate of our
model.

Given these uncertainties, we do not simulate only our {\small\sc fiducial} dust model but also study model variations, which are summarised in
Table~\ref{table:dust_models}. We specifically explore variations of the
thermal sputtering rate, the dust IR cooling rate, and the dust grain size.
Thermal sputtering is the main mechanism that destroys dust in the ICM, and we
expect that variations in the sputtering timescales significantly affect the
amount of dust in the ICM. There are additional uncertainties in the dust
cooling rates, and we therefore also explore a model where we increase the dust
cooling rate.  This is mainly to explore how strongly dust cooling can actually
affect the thermodynamic state of the cluster gas in the ICM. Our model also
assumes a single grain size, while the actual dust population follows a certain
grain size distribution, so we explore also variations of the grain
size.  Unlike the changes in sputtering and cooling rates, the change in grain
size affects multiple dust processes since this quantity enters the growth
rates, the sputtering timescales and the cooling rates. We expect that these
model variations roughly bracket the overall uncertainties of our model. In the following
Section we will explore these model variations to understand the abundance, distribution, and
impact of dust in the cluster environment.

\section{Global Dust content}
\label{sec:Section3}

To get a first impression of the simulation results we present in
Figure~\ref{fig:dust_maps_1} dust maps of the cluster at $z=0$ for the
different dust models presented in Table~\ref{table:dust_models}. The top row
shows the dust surface density, the middle row the metal surface density, and
the bottom row the dust-to-metal ratio. The maps in the three rows are related
since dust production occurs at the expense of gas phase metals.
Furthermore dust destruction in the ICM due to thermal sputtering returns
metals to the gas phase. The dust maps reveal that the amount of dust in the
ICM increases significantly for the model with reduced thermal sputtering,
{\small\sc slow-sputter}.  This is expected given that thermal sputtering is
the main destruction channel for dust in the ICM, where SNII destruction cannot
occur. We can also identify the main production sites of dust as the
cluster member galaxies, where dust is produced during the stellar evolution
process and grows in the ISM. Stripping of gas, metals and dust from these
cluster member galaxies then enriches the ICM. The overall amount of dust in
the ICM is set by the competition between stripping and thermal sputtering. The
stripping of dust is also visible in the dust maps through various stream like
features. The model with slow thermal sputtering, {\small\sc slow-sputter},
shows a rather uniform distribution of dust in the ICM. This is reminiscent of
the uniform distribution of metals in the ICM that is both found
observationally and in simulation studies~\citep[e.g.,][]{Werner2013, Vogelsberger2018}.
For these low sputtering rates, the dust component behaves similar to the gas
phase metals in the ICM since it is stripped from member galaxies and then
mixes within the ICM but is not destroyed due to the lack of efficient thermal
sputtering. Furthermore, dust in the ICM does not experience any other growth
or destruction processes. The fact that dust traces the metal distribution very
well for the {\small\sc slow-sputter} model can also be seen in the
dust-to-metal ratio map for this model. This map is nearly constant except for
deviations in the inner region, which are caused by dust growth and destruction
in the ISM of the central galaxy. For all other dust models, the dust-to-metal
ratio maps show fluctuations in the cluster gas, which are caused by the non-negligible
thermal sputtering occurring for these models. For example, for the {\small\sc
fiducial} model we can see quite large fluctuations in the dust-to-metal ratio
within the cluster. These fluctuations can be as large as two orders of
magnitude.  We can also see that the model variations with more cooling,
{\small\sc more-cool}, or larger grains, {\small\sc large-grains}, produce
slightly more dust in the cluster. The increased cooling leads to an increased
production of dust, and a larger grain size leads to a lower thermal sputtering
rate since the timescale for thermal sputtering depends linearly on the dust
grain size for a single grain size population as adopted in our study.  We note
that a larger grain size will also slow down the dust growth in the ISM of
galaxies and at the same time also reduce the dust IR cooling. However, the
impact of thermal sputtering on the total amount of dust in the cluster is larger
than these other two effects, and we therefore see an overall increase of the
dust mass for larger grains. These results are consistent with the
findings of~\cite{Gjergo2018}, who also found that large grains are more
abundant in the cluster gas since smaller ones are destroyed more quickly due to
sputtering. We also note that the maximum dust surface densities in the
{\small\sc fiducial} models reach values of about $\sim 0.1\msun\pc^{-2}$,
which is roughly in agreement with the findings in~\cite{Gjergo2018}, and also
consistent with observational constraints as we will discuss below.

\begin{figure}
\centering
\includegraphics[width=0.475\textwidth]{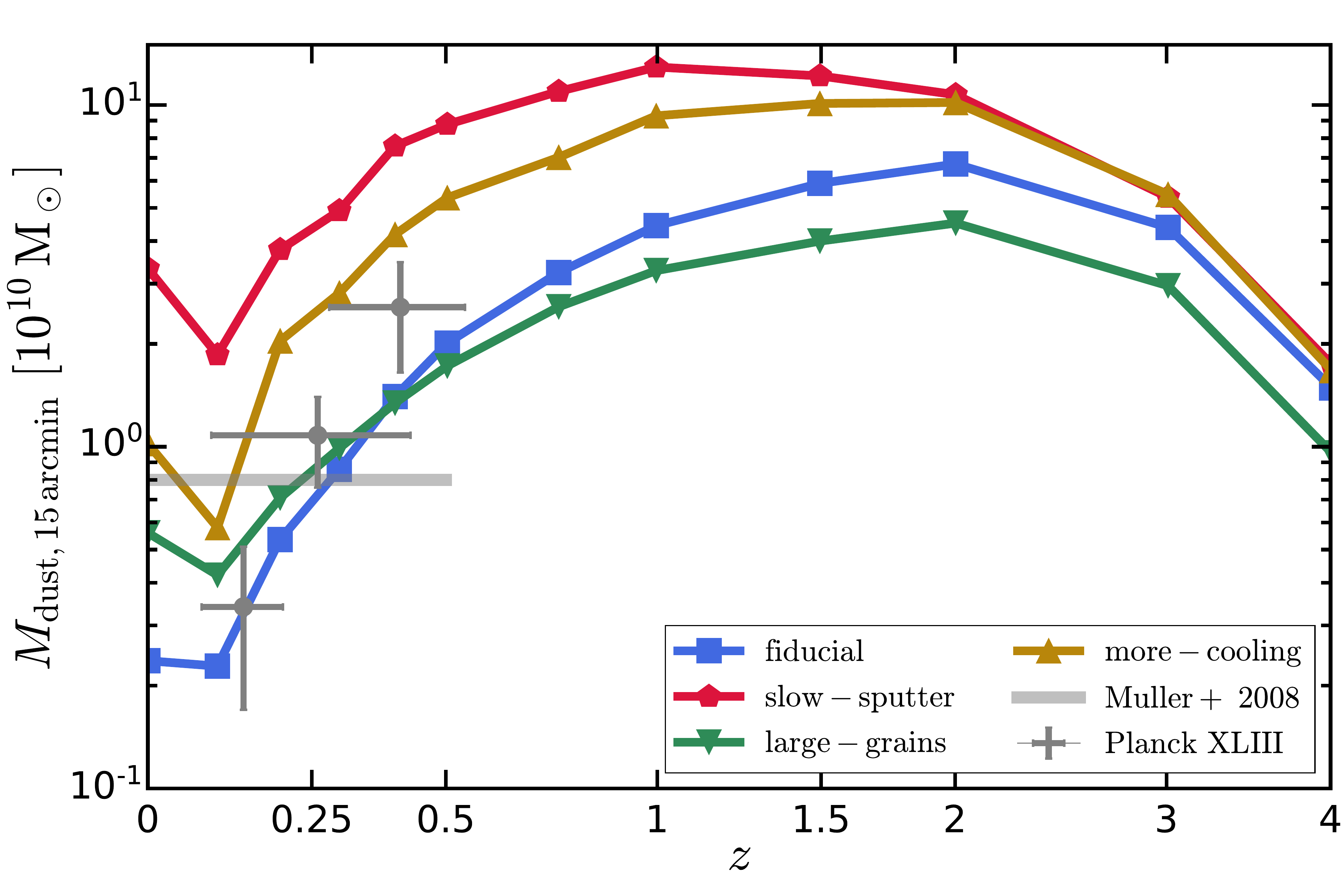}
\caption{{\bf Redshift evolution of the total dust mass.} We measure the total
dust mass at each redshift within $15\,{\rm arcmins}$. At lower redshift, once
$15\,{\rm arcmins}$ is smaller than $r_{\rm 200, crit}$, we replace this radial
cut with $r_{\rm 200, crit}$. The qualitative behaviour of the dust mass
evolution is the same for all models, and the dust mass typically peaks around
$z\sim 1.5 - 2$. The dust mass then starts to decline again reaching a minimum
at the present-day. This shape is determined by two factors: First, dust
production is related to star formation, which peaks at around $z\sim 2$. And
second thermal sputtering, which destroys dust very efficiently in the ICM, is most active
towards lower redshifts once the cluster ICM gas reaches high temperatures. We also include
observational dust estimates from the \protect\cite{PlanckCollaboration2016}
and \protect\cite{Muller2008}. Our {\small\sc fiducial} model is marginally
consistent with this data, but the {\small\sc more-cool} dust model results in
a slightly better agreement, especially for the highest redshift observational
data point.}
\label{fig:dust_time_evolution_2}
\end{figure}

Next we study the time evolution of the total amount of dust in the cluster
more quantitatively.  These results are presented in
Figure~\ref{fig:dust_time_evolution_1}, where we show the dust mass evolution
as a function of redshift. The left panel shows the total dust mass within
$15\,{\rm arcmins}$ around the cluster center at each redshift. This specific radial cut is chosen
to compare to observational constraints from Planck as described
below. At lower redshifts we replace this radius with a restriction within
$r_{\rm 200,crit}$ once the radius corresponding to $15\,{\rm arcmins}$ is
smaller than $r_{\rm 200, crit}$. We show the enclosed dust mass relative to
the total cluster mass within $r_{\rm 200,crit}$, i.e. $M_{\rm 200,crit}$. We
find that this ratio decreases towards lower redshifts, and drops for most
models by more than three orders of magnitude from $z=4$ to $z=0$. Around a
redshift of $\sim 4$ the different models predict nearly the same amount of
dust roughly corresponding to $\sim 1\%$ of the total cluster mass being in
dust. This ratio decreases then for the {\small\sc fiducial} model towards
$z=0$, where we predict that the total amount of dust in the cluster is only
$\sim 3 \times 10^{-6}$ of $M_{\rm 200, crit}$. This decrease of dust is
largely driven by dust destruction in the ICM due to thermal sputtering. For
example, the model with slower sputtering, {\small\sc slow-sputter}, has a dust
mass which is about one order of magnitude larger than the dust mass of the
{\small\sc fiducial} model at $z=0$. Switching to larger grains also increases
the dust mass compared to the {\small\sc fiducial} model since the larger
grains lead to slower sputtering due to the linear dependence of the sputtering
timescale on the grain size as discussed above. The model with increased
dust cooling rates also leads to a larger dust mass caused by more dust
production due to the increase in the overall cooling rate. 

The right panel of Figure~\ref{fig:dust_time_evolution_1} also shows the total
dust mass within $15\,{\rm arcmins}$ but divided by the total gas mass within
$r_{\rm 200,crit}$. Again we replace $15\,{\rm arcmins}$  by $r_{\rm 200,crit}$
at lower redshifts. The panel essentially shows the time evolution of the
average dust-to-gas ratio within the cluster.  This ratio also decreases 
towards lower redshifts.  The {\small\sc fiducial} model predicts a global
dust-to-gas ratio of about $\sim 2 \times 10^{-5}$ at $z=0$.  The model with
strongly reduced sputtering, {\small\sc slow-sputter}, on the other hand,
predicts a dust-to-gas ratio about one order of magnitude larger at $z=0$. The
ordering for the different models at various redshifts is the same in the two
panels of Figure~\ref{fig:dust_time_evolution_1}, demonstrating that the
cluster gas fractions between the different dust models must be rather similar.
We have confirmed explicitly that the gas fractions indeed only vary at the
percent level between the different dust models.

So far we have not yet confronted our dust predictions with observations,
which is a crucial test for the validity of our dust model.  Unfortunately,
little is known observationally about the dust content of galaxy clusters and their ICM due the
difficulties of its detection both through IR emission or reddening. Here we
compare our results to the cluster dust IR measurements of the
\cite{PlanckCollaboration2016}.  Although IR dust emission from clusters of
galaxies had already been statistically detected using IRAS data,
it has not been possible to sample the spectral energy distribution of this
emission over its peak, which is required to break the degeneracy between dust
temperature and mass. The \cite{PlanckCollaboration2016} provided new
constraints on the IR spectrum of thermal dust emission in clusters of galaxies
improving on these existing cluster IR detections. We include in both panels of Figure~\ref{fig:dust_time_evolution_1} dust
mass estimates from Planck based on their full cluster sample with an average
mass of ${M_{\rm 200,crit} = (5.6 \pm 2.1) \times 10^{14} \, {\rm M}_\odot}$.
In addition we also include the low redshift ($z < 0.25$) and high redshift
($z>0.25$) Planck samples with average masses of ${M_{\rm 200,crit} = (4.3 \pm
1.7) \times 10^{14} \,{\rm M}_\odot}$ and ${M_{\rm 200,crit} = (7.0 \pm 1.5)
\times 10^{14} \,{\rm M}_\odot}$, respectively. We note that other
observational studies found similar dust mass fractions. For example,
\cite{Guti2017} reported, based on $327$ clusters in the redshift range $0.06
- 0.7$, that dust should contribute a fraction of about $9.5 \times 10^{-6}$
  to the total cluster mass. This value, however, refers only to dust in the ICM
since they subtracted known dust contributions from cluster galaxies. 
The \cite{PlanckCollaboration2016} also
provided observational estimates for the total dust-to-gas ratios; i.e. the
ratio of the total dust mass measured within $15\,{\rm arcmins}$ over the gas
mass contained within $r_{\rm 200,crit}$, where they assumed ${M_{\rm gas, 200}
\cong 0.1 \times M_{\rm 200, crit}}$.  We include those estimates in the right
panel of Figure~\ref{fig:dust_time_evolution_1}. We caution here that the Planck
cluster samples have different average masses than the halo we study here at the
various redshifts.
Specifically, at ${z=0.26 \pm 0.17}$, which is the average redshift of the full
Planck sample, we find that ${M_{\rm 200,crit}=3.8 \times 10^{14}\,{\rm M}_\odot}$ at
$z=0.3$ for the simulated cluster in this work, which is therefore slightly
less massive than the average cluster in the Planck sample.  The low redshift
Planck sample has an average redshift of ${z=0.139 \pm 0.063}$; and at $z=0.1$
the simulated cluster has a mass of ${M_{\rm 200,crit}=4.6 \times 10^{14}\,{\rm
M}_\odot}$, which is quite close to the average mass of the Planck low
redshift sample. The high redshift Planck sample has an average redshift of
${z=0.41 \pm 0.13}$; and at $z=0.4$ the simulated cluster has a mass of
${M_{\rm 200,crit}=3.4 \times 10^{14}\,{\rm M}_\odot}$, which is less massive
than the average mass of the Planck high redshift sample. 

With all these caveats in mind, we can compare the different dust models to the
Planck data in both panels of Figure~\ref{fig:dust_time_evolution_1}. We find
that this observational data agrees reasonably well with our {\small\sc
fiducial} model, which has a reduced thermal sputtering rate compared to the
original \cite{McKinnon2017} model as mentioned above. We also find that the
model with larger grains is consistent with the Planck data as well. The model with increased dust cooling rates is slightly inconsistent producing
too much dust compared to the Planck data. The {\small\sc slow-sputter} model, however,
overpredicts the amount of dust significantly as can be seen in both panels of
Figure~\ref{fig:dust_time_evolution_1}. Knowing that the {\small\sc fiducial}
model describes the low redshift observational data correctly, we can also make
predictions for the cluster dust content at higher redshifts using this model.
For example, at $z=1$ our {\small\sc fiducial} model predicts a dust mass
fraction of around $\sim 10^{-4}$ and a dust-to-gas ratio of around $10^{-3}$.
Finally, we note that the dust-to-gas ratio values of our {\small\sc fiducial}
model is consistent with other observational findings. For example,
\cite{Roncarelli2010} studied the IR emission of clusters with $0.1 < z < 0.3$
found an upper limit of $\lesssim 5 \times 10^{-5}$.

\begin{figure}
\centering
\includegraphics[width=0.475\textwidth]{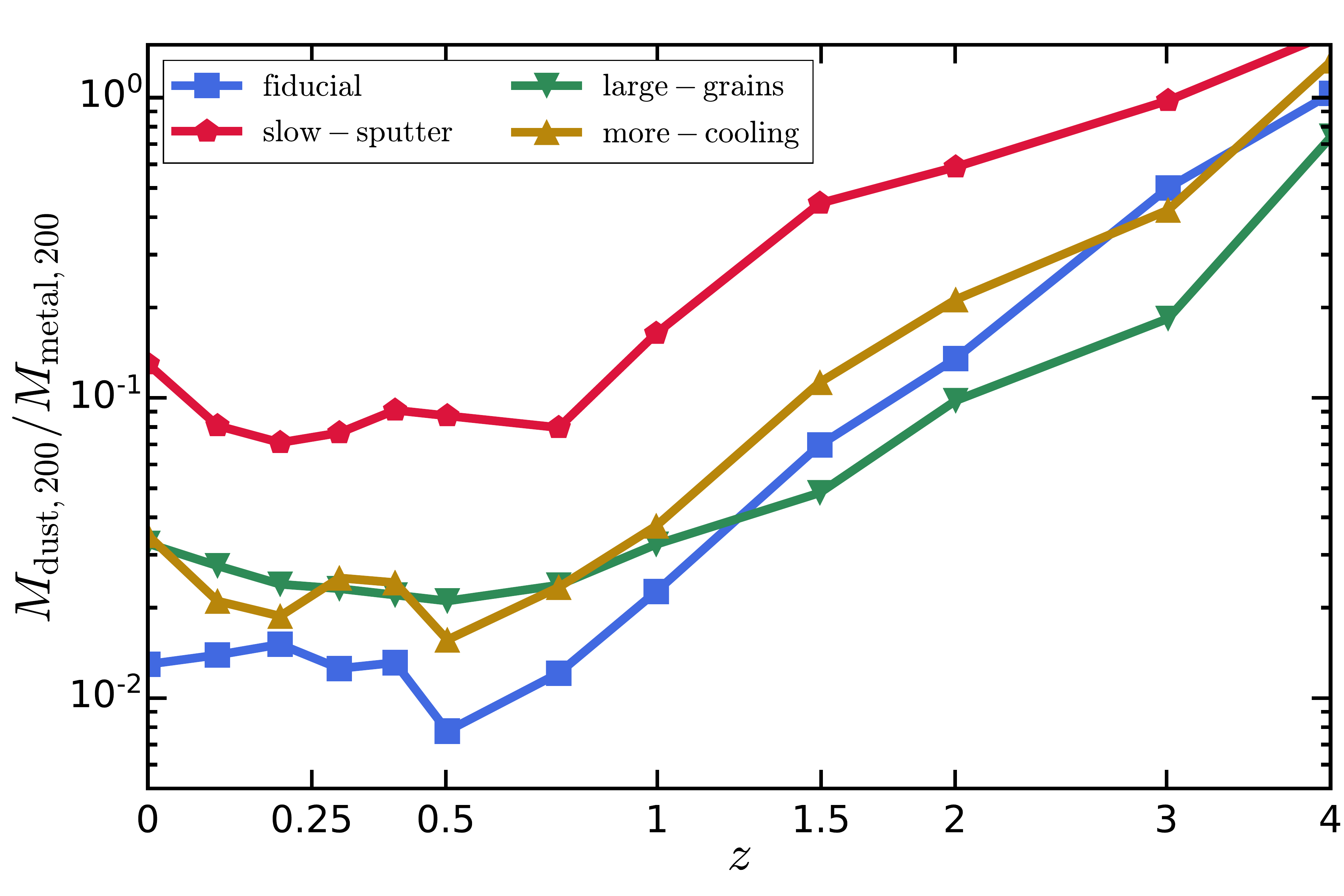}
\caption{{\bf Redshift evolution of the dust-to-metal-ratio.} We measure the
dust and gas phase metal mass at each redshift within $r_{\rm 200, crit}$ and
plot the ratio as a function of redshift. The high redshift behaviour of the
different models is rather similar, except for the {\small\sc slow-sputter}
model, which has a higher dust-to-metal ratio already at higher redshifts. The
other variations of the {\small\sc fiducial} model lead to rather similar
dust-to-metal ratios. The {\small\sc fiducial} model predicts at $z=0$ a
dust-to-metal ratio of about $\sim 10^{-2}$. The {\small\sc large-grains} model
leads to a slightly higher dust-to-metal ratio of about $\sim 4 \times
10^{-2}$. The model with reduced thermal sputtering, {\small\sc slow-sputter} leads to
a ten times larger dust-to-metal ratio compared to the {\small\sc fiducial} model. }
\label{fig:dust_time_evolution_3}
\end{figure}

Besides comparing to observational data, we can also compare our predictions to
the recent cluster dust simulations of~\cite{Gjergo2018}. Their model shares
similarities with ours, but also some differences. For example, they consider
two grain populations, large and small, but do not include dust IR cooling.
The implementation of the growth timescales also differs between the two
models. Furthermore, the underlying galaxy formation model and simulation methods 
are also rather different. In the left panel of
Figure~\ref{fig:dust_time_evolution_1} we show the dust mass redshift evolution
of their halo with a redshift zero mass of $5.4\times 10^{14}\msun$ in their
reduced sputtering model with a five times larger sputtering timescale compared
to their fiducial model. The $z=0$ cluster mass of their cluster is nearly
identical to our cluster mass, $5.4\times 10^{14}\msun$.  Interestingly,
\cite{Gjergo2018} found that they also had to increase their fiducial
sputtering timescale to be consistent with observational data for the total
dust mass. We can compare their effective sputtering timescale with ours, and
find that for a given density, temperature and grain size their best-fit
thermal sputtering timescales is nearly identical with ours. Specifically, our
{\small\sc fiducial} model thermal sputtering timescale is only about $5\%$
larger than the one presented in~\cite{Gjergo2018}. Despite the fact that the
details of the implementation of the other dust model processes and the galaxy
formation model do vary between the models, it seems that both models favour a
nearly identical thermal sputtering timescale.  We note that the adopted
density and temperature dependence of the sputtering timescale is also
identical between the two models. It therefore seems that, at least based on
these two simulation results, dust can survive longer in the ICM than expected
based on the fiducial model of \cite{Gjergo2018} and the original model of
\cite{McKinnon2017}. This seems to be a rather robust prediction given that
both models vary in most other model parts. We note that the two model
predictions deviate a bit towards higher redshifts, and that the normalisation
towards lower redshift is a bit smaller for the~\cite{Gjergo2018} simulations
compared to our {\small\sc fiducial} model predictions.  Nevertheless, the
overall redshift evolution and the present-day predictions are in remarkable
agreement between the two models.  This is most likely related to the fact that
thermal sputtering is the most important process for setting the overall dust
budget in the ICM.

Instead of showing ratios of the dust mass with respect to the total halo mass
or gas mass, we can also inspect the absolute total dust mass within the
cluster and its time evolution.  In Figure~\ref{fig:dust_time_evolution_2} we
plot the total dust mass as a function of time within $15\,{\rm arcmins}$
without normalising to the total cluster mass or gas mass. As discussed above
we replace this radial cut with $r_{\rm 200,crit}$ at lower redshifts.  We also
added the dust mass estimates from the \cite{PlanckCollaboration2016} for the
full, low redshift, and high redshift sample to that plot. For the whole sample,
Planck obtained an average dust mass of ${(1.08 \pm 0.32) \times 10^{10}\,{\rm
M}_\odot}$. This estimate is similar to the values obtained with different
techniques.  For example, \cite{Muller2008} found ${8 \times 10^{9}\,{\rm
M}_\odot}$ for a sample of comparable redshift distribution. For a relatively
low mass sample \cite{Gutierrez2014} found dust masses ${< 8.4 \times
10^{9}\,{\rm M}_\odot}$. \cite{Guti2017} estimated the typical dust mass of
their sample to be $\sim 2\times 10^9\msun$.  Planck finds for their low
redshift sample ${(0.34 \pm 0.17) \times 10^{10}\,{\rm M}_\odot}$, and ${(2.56
\pm 0.91) \times 10^{10}\,{\rm M}_\odot}$ for the high redshift sample. The
difference in dust mass between these two samples is largely driven by the
different average masses of the two samples. 
We note that the dust mass
redshift evolution revealed by Figure~\ref{fig:dust_time_evolution_2} shows an
increase towards $z \sim 1.5$ and then a decrease. The peak roughly occurs at
the time of maximum star formation rate activity when dust production is
largest. After that the amount of dust decreases due to thermal sputtering
in the hot atmospheres of the cluster.  Comparing the observational estimates
with the different dust models, we can see that the {\small\sc fiducial} model
is within the error bars for at least the lowest two redshift points. However,
the higher redshift data point has a dust mass above the mass predicted from
the {\small\sc fiducial} model for that redshift. The {\small\sc
more-cooling} model results in a slightly too high dust mass that compared to the observational
data. Combining these findings with the time evolution shown in
Figure~\ref{fig:dust_time_evolution_1} we conclude that the {\small\sc
fiducial}, {\small\sc large-grains}  and {\small\sc more-cooling} are roughly consistent with the observationally
inferred cluster dust masses.

The production of dust goes along with the production of gas phase metals since
dust essentially consists of condensed gas phase metals. We therefore show in
Figure~\ref{fig:dust_time_evolution_3} the time evolution of the dust-to-metal
ratios within $r_{\rm 200, crit}$. This ratio also decreases towards lower
redshifts, which is again driven by thermal sputtering,
efficiently destroying dust grains in the hot ICM, and correspondingly
increases the abundance of metals in the ICM. The importance of thermal
sputtering can also be inferred by comparing the {\small\sc fiducial} model to
the model with lower sputtering rates, {\small\sc slow-sputter}. This model
shows a slower decrease of the dust-to-metal ratio towards lower redshifts, and
the cluster ends up with a larger amount of dust at $z=0$ for this model. Our
{\small\sc fiducial} model predicts at $z=0$ a dust-to-metal ratio of close to
$10^{-2}$, which is consistent with the dust-to-gas ratios measured by Planck
assuming an average cluster metallicity of $\sim 0.3$ solar. Specifically,
for the lowest redshift Planck dust-to-gas ratio we can infer a dust-to-metal
ratio of $\sim 2 \times 10^{-2}$ assuming a solar metallicity of $Z_\odot =
0.0127$. This value is roughly consistent with the prediction of our {\small\sc
fiducial} model. The simulated dust-to-metal ratios stay rather constant back
to $z\sim 1$ and then increase towards higher redshifts. At $z=2$ we find a
ratio of $\sim 0.1$ for the {\small\sc fiducial} model. At even higher
redshifts, $z\sim 4$, we find nearly the same amount of metals and dust within
the cluster.  This high redshift prediction is rather independent of the model
variations since the main physical process that affects the dust content of the
cluster, thermal sputtering, is not yet active at these high redshifts because
of a too low gas temperature in the ICM. A common trend seen for all models
is that the dust-to-metal ratio stays rather constant starting at
roughly $z \sim 1$ all the way down to $z=0$. The normalisation of the ratio
differs between the models.

We conclude from this Section that our {\small\sc fiducial} dust model agrees
well with current observational data and makes predictions for the
high redshift dust content of the clusters and their ICM. Furthermore, our results also agree
with the recent simulations of~\cite{Gjergo2018}.

\section{Dust Profiles}
\label{sec:Section4}

Our simulation not only predicts the total amount of dust of the cluster
but also the spatially resolved dust distribution. We can therefore construct
various radial dust profiles to understand not only the overall abundance of
dust in the cluster, but also its spatial distribution. In this Section we will
explore these dust profiles.

\begin{figure}
\centering
\includegraphics[width=0.495\textwidth]{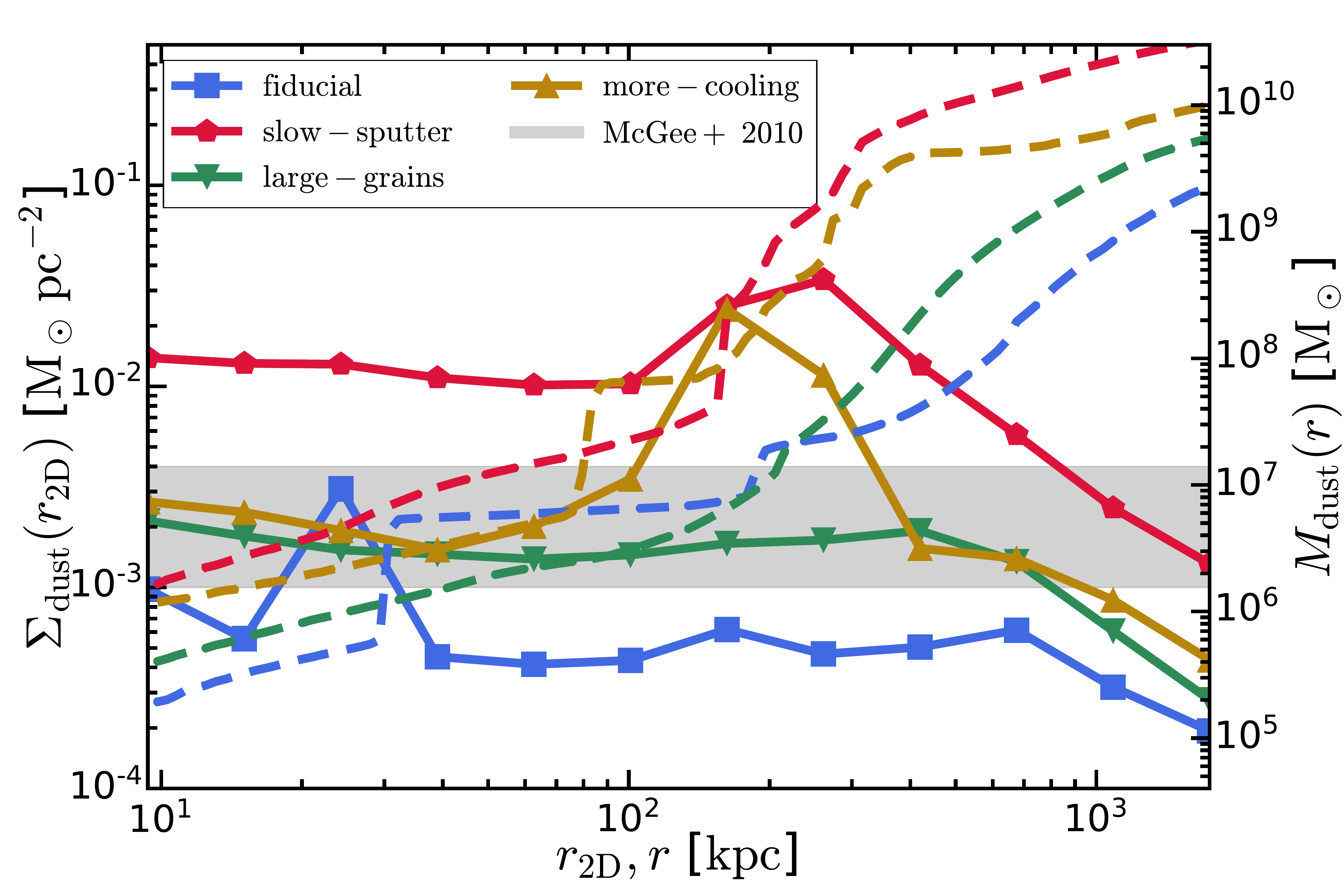}
\caption{{\bf Dust surface density and cumulative dust mass profiles at
$\mathbf{z=0}$.} The solid lines show the dust surface density profiles as a
function of the two-dimensional radius, $r_{\rm 2D}$.  The dashed lines show
the corresponding cumulative dust mass profiles as a function of the
three-dimensional radius, $r$. The observational band is taken from
\protect\cite{McGee2010} showing some estimate for the observationally
expected dust surface density averaged over all cluster gas. The {\small\sc
fiducial} model is consistent with these observational values, but the
{\small\sc slow-sputter} model overshoots them. All models predict very flat
dust surface density profiles over the full radial range. This is also consistent
with observational findings. }
\label{fig:dust_surface_density}
\end{figure}

\begin{figure*}
\centering
\begin{tabular}{ll}
\includegraphics[width=0.49\textwidth]{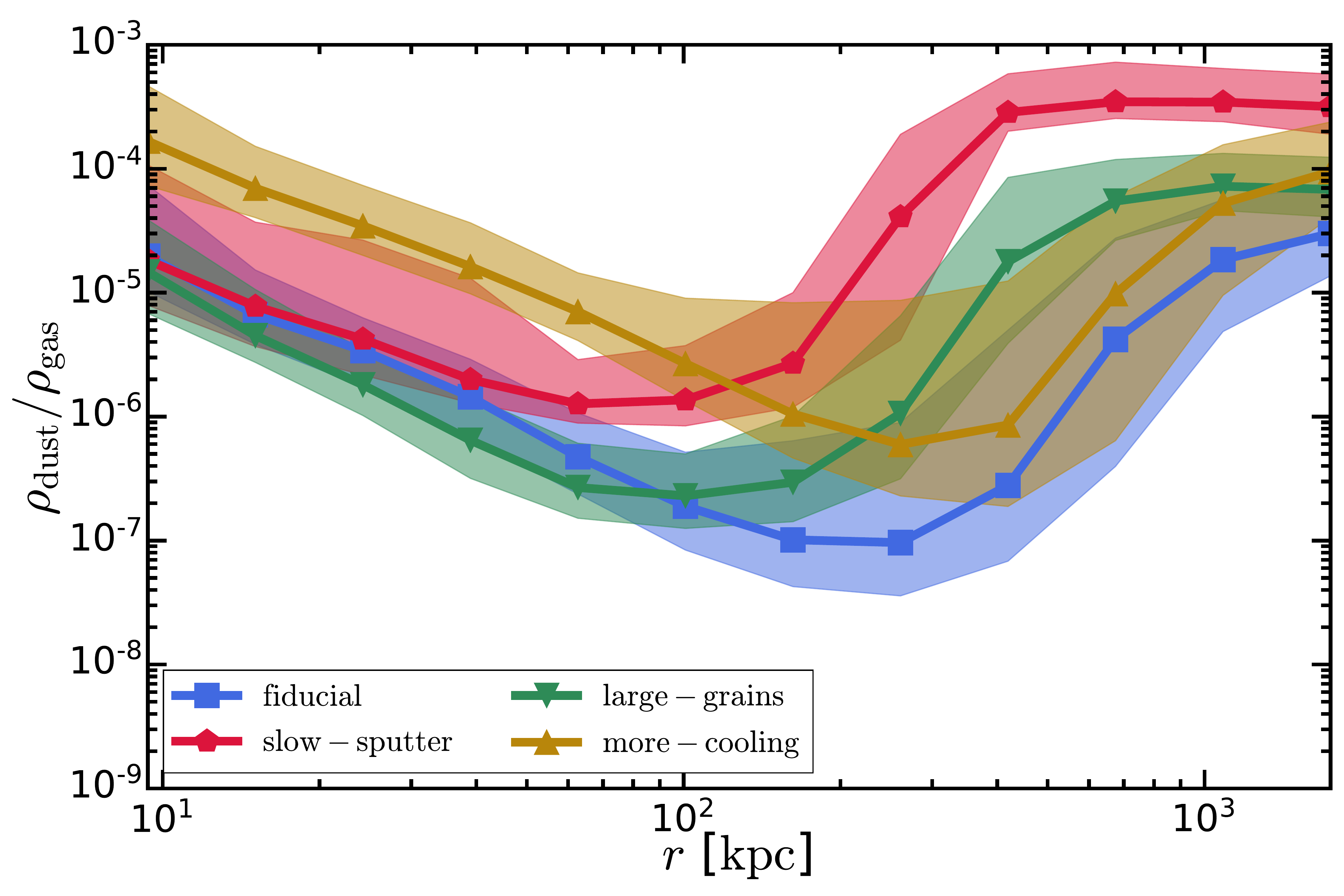}
\includegraphics[width=0.49\textwidth]{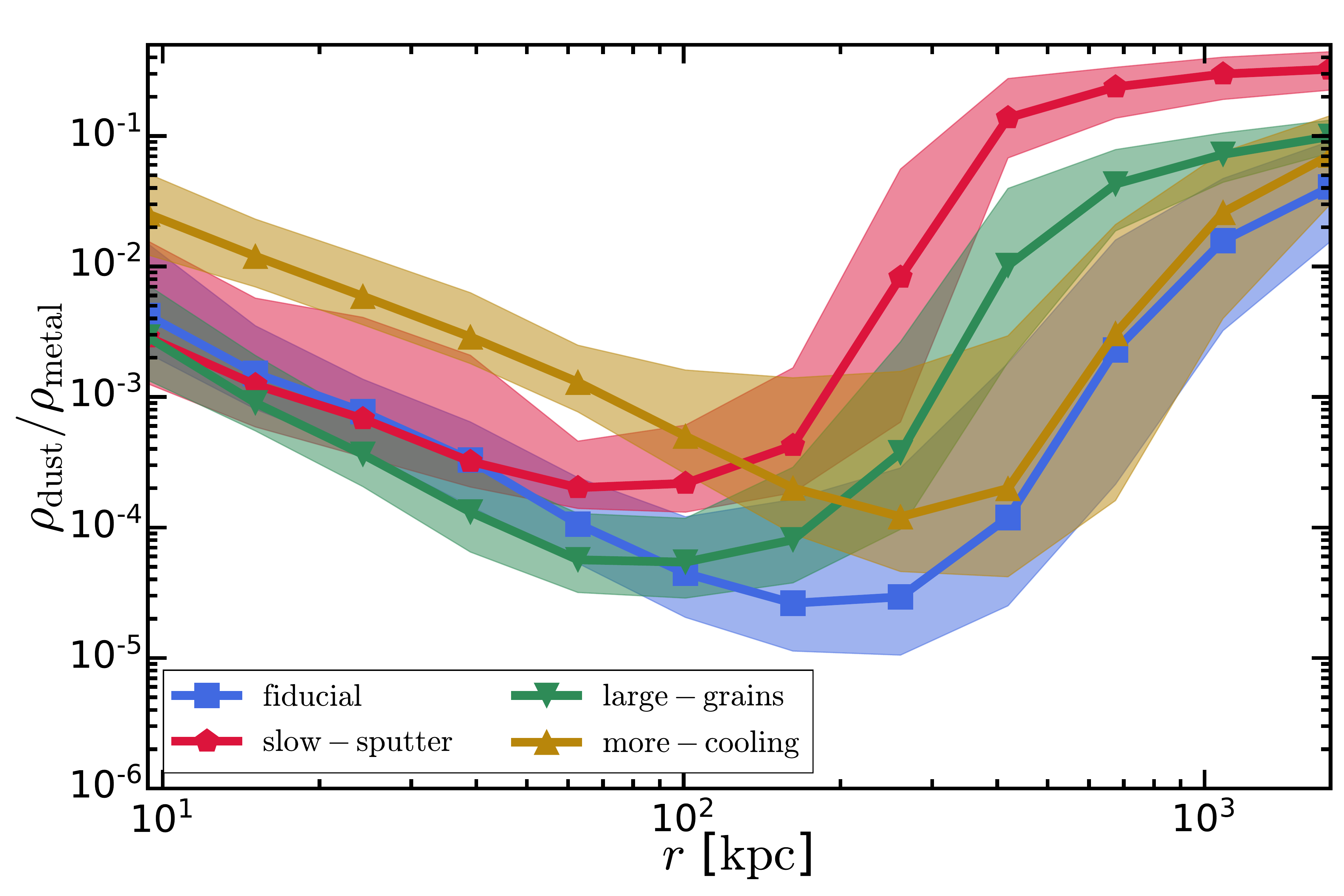}
\end{tabular}
\caption{{Radial dust-to-gas and dust-to-metal profiles at $\mathbf{z=0}$.}
{\it Left panel:} Dust-to-gas radial profiles. {\it Right panel:} Dust-to-metal
radial profiles. The profiles show the median and $1\sigma$ scatter around that
median.  All profiles follow a rather similar qualitative shape with ratio
minima around $\sim 100-500\kpc$ and increasing ratios towards the center and
outskirts of the ICM. These shapes can be explained by the behaviour of the
thermal sputtering timescales and the fact that dust production and growth is
strongest towards the central regions of the cluster.}
\label{fig:dust_to_gas}
\end{figure*}

We begin with Figure~\ref{fig:dust_surface_density}, which shows dust surface
density profiles (solid lines), and cumulative dust mass profiles (dashed
lines).  We note that the dust surface density is plotted as a function of a
two-dimensional radius, $r_{\rm 2D}$, whereas the cumulative dust mass is shown
as a function of the spherical radius, $r$. Consistent with the results for the
time evolution of the dust mass, we find that the model with reduced
sputtering rates, {\small\sc slow-sputter}, leads to a larger dust surface density
and cumulative mass. Specifically, the model with slow sputtering leads to a
total dust mass greater than $10^{10}\msun$ within $1\Mpc$. Furthermore, this
model also predicts a high dust surface density within a few hundred ${\rm
kpc}$ of about $\sim 10^{-2}\msun\pc^{-2}$.  The gray band shows some
observational estimate for the dust surface density taken from the innermost
radial bin of the dust surface density profile presented in~\cite{McGee2010}
based on a reddening study of groups and clusters. The {\small\sc
slow-sputter} model overshoots this observational data substantially. The
{\small\sc fiducial} model, which agrees well with the total dust mass
estimates as demonstrated in the previous Section, leads to a slightly too low
dust surface density. We note however, that these observational estimates of dust
surface densities  have rather large systematic uncertainties, and
we therefore conclude that except for the slow sputtering model, all
our models are likely consistent with these observed dust surface density values. Indeed, 
our surface density predictions are also consistent with the few
other observational estimates for the dust surface density in clusters. For
example, \cite{Kitayama2009} studied the IR emission of the Coma cluster and
found an upper limit of $1.4 \times 10^{-3}\msun\pc^{-2}$ by combining their
measurements with the theoretical model of \cite{Yamada2005}. 

The profiles of the dust surface density demonstrate that all dust models
show a rather flat dust surface density profile within $1\Mpc$. This therefore
seems to be a rather generic result. For example, the 
predicted dust surface density of the {\small\sc fiducial} changes only very little going from beyond $1\Mpc$ to the inner $10\kpc$; except for the inner spike. 
Interestingly, the reddening study from \cite{McGee2010} based on $70\,000$
uniformly selected galaxy groups and clusters also found evidence for
a relatively uniform distribution of dust in clusters, which is in qualitative
agreement with our results.  The cumulative dust mass profiles follow the same
order as the $z=0$ data in Figure~\ref{fig:dust_time_evolution_2}. For most of
the models this order also stays the same at different radii.

Dust abundances are often quantified as dust-to-gas and dust-to-metal ratios as
discussed above. We therefore show in Figure~\ref{fig:dust_to_gas}
the median profiles for dust-to-gas (left panel) and dust-to-metal (right
panel) ratios. The shaded regions mark the $1\sigma$ spread around those
medians. The different models produce rather similar functional shapes for the
dust-to-gas and dust-to-metal profiles, but with quite different
normalisations. Specifically, these profiles have a minimum at intermediate
radii around $\sim 100-200\kpc$ and increase towards smaller and larger
cluster-centric distances.  For the dust-to-gas ratio we find that the
difference between this minimum and the larger values in the inner and outer
parts of the cluster decreases for the model with reduced sputtering.  The same
trend can be seen for the dust-to-metal ratios. The reason for this behaviour
is that the model with less sputtering leads to an overall more uniform and
higher level of dust in the ICM. In fact, we have tested that a model with no
thermal sputtering at all leads to a nearly flat profile for both ratios
because no dust is being destroyed in this case.  For the {\small\sc fiducial}
model we find on the other hand that the lowest dust-to-total-mass ratio is around
$\sim 10^{-7}$ and occurs at radii slightly larger than $\sim 100\kpc$. The largest
ratio occurs in the outer parts of the cluster, where we find a
ratio of the order of $\sim 3\times 10^{-5}$, which is slightly larger than the
central value of $\sim 10^{-5}$. We find a similar trend for the dust-to-metal
ratios in the right panel. Here the {\small\sc fiducial} model predicts a central
dust-to-metal ratio of about $\sim 3\times 10^{-3}$, while the value in the
outer cluster parts increase to ratios larger than a few times $\sim 10^{-2}$.

\begin{figure}
\centering
\includegraphics[width=0.475\textwidth]{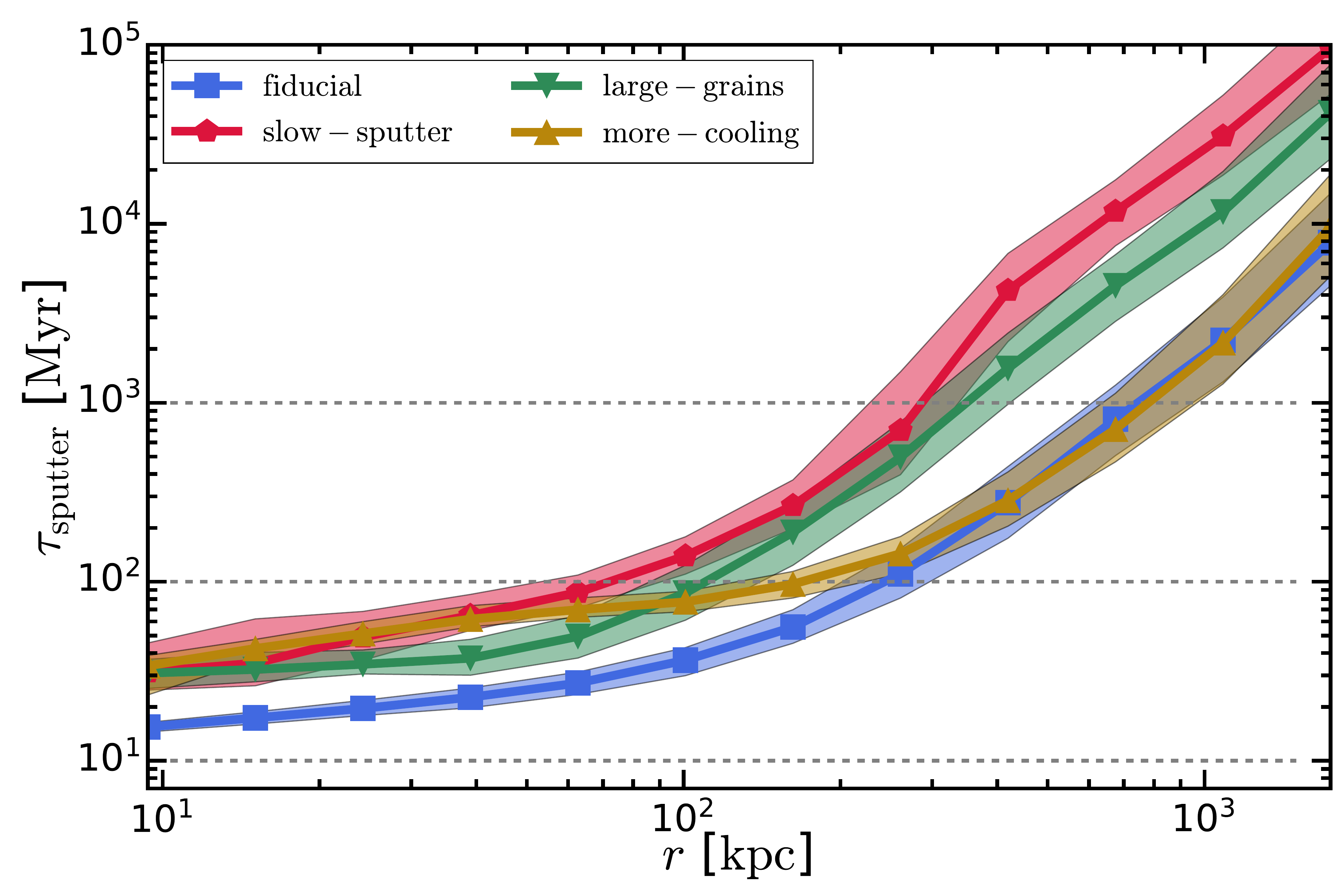}
\caption{{\bf Radial profile of the sputtering timescale for the different dust
models at $\mathbf{z=0}$.}  The profiles show the median and $1\sigma$ scatter
around that median. Horizontal thin lines mark three different timescales for
orientation.  The sputtering timescales of the {\small\sc large-grains}
simulation are nearly as long as those of the {\small\sc slow-sputter} model
due to the grain size dependence of the sputtering timescale. The thermal
sputtering profile is set by the underlying gas temperature and gas density
profiles of the cluster gas such that the sputtering timescales are shortest in
the center of the cluster and longest in its outskirts. Our {\small\sc
fiducial} model predicts sputtering timescales of $\sim 10\Myr$ in the inner
parts of the ICM, and about $\sim 1\Gyr$ towards the outer parts. }
\label{fig:tausputter}
\end{figure}

To qualitatively understand the origin of the particular functional shape of
both of these ratio profiles, we have to inspect primarily the thermal sputtering
behaviour in the ICM since this is the main dust process affecting
the dust abundance in the cluster. The efficiency of thermal sputtering is set by the
thermal sputtering timescale $\tau_{\rm sputter}$. In Figure~\ref{fig:tausputter}
we therefore show radial profiles of the median thermal sputtering timescales for the
different dust models. The shaded regions show the $1\sigma$ scatter around
these profiles. We consider here only non-star-forming gas in the ICM. All thermal sputtering timescale
profiles have a quite similar shape, where timescales are shortest in the center
and longest in the outer part of the halo. In addition the spread around those
median values is relatively small. The thermal sputtering timescale
depends on some of the dust model parameters, most importantly the parameters
which directly influence the overall sputtering timescale normalisation.  This is obvious for the {\small\sc
slow-sputter} model, where the $\tau_{\rm sputter}^{\rm ref}$ value has been
increased by factors of $10$ with respect to the {\small\sc fiducial} model.
Similarly a change in grain size also directly changes the overall
normalisation of the sputtering timescale as it depends linearly on the
size of dust grain. Therefore, the model with five times larger grains, {\small\sc large-grains}, will typically have five times larger sputtering
timescales. For all models, we find that despite the variations in normalisation,
the sputtering timescales are shorter than $\sim 100\,{\rm
Myr}$ within $\sim 100\kpc$. Beyond a $\sim 1\Mpc$ distance from the center all models
predict also sputtering timescales which are larger than $\sim 1\,{\rm Gyr}$.
The {\small\sc slow-sputter} model predicts timescales even longer than $\sim
10\,{\rm Gyr}$ in these outer regions of the ICM. Any dust in this part of the
ICM will therefore be able to survive very long times without being destroyed through
sputtering.

The overall shape of the sputtering profiles is set through the underlying gas
density and temperature profiles, which regulate the thermal sputtering
timescale.  Gas densities and temperatures are highest towards the central part
of the cluster, and therefore sputtering is most efficient in this region of
the ICM, while the opposite is true for the outer parts of the halo. This explains
the general shape of these thermal sputtering profiles. We note that both the
thermal sputtering timescale profiles and the entropy profiles of clusters are
functions of temperature and density. It is therefore not too surprising that
they share some similarities like different functional forms in the inner and
outer part. In fact, we will see below that the gas entropy profiles
also change functional shape at a radius of $\sim 100-200\kpc$ as the
thermal sputtering profiles do.

\begin{figure*}
\centering
\begin{tabular}{ll}
\includegraphics[width=0.49\textwidth]{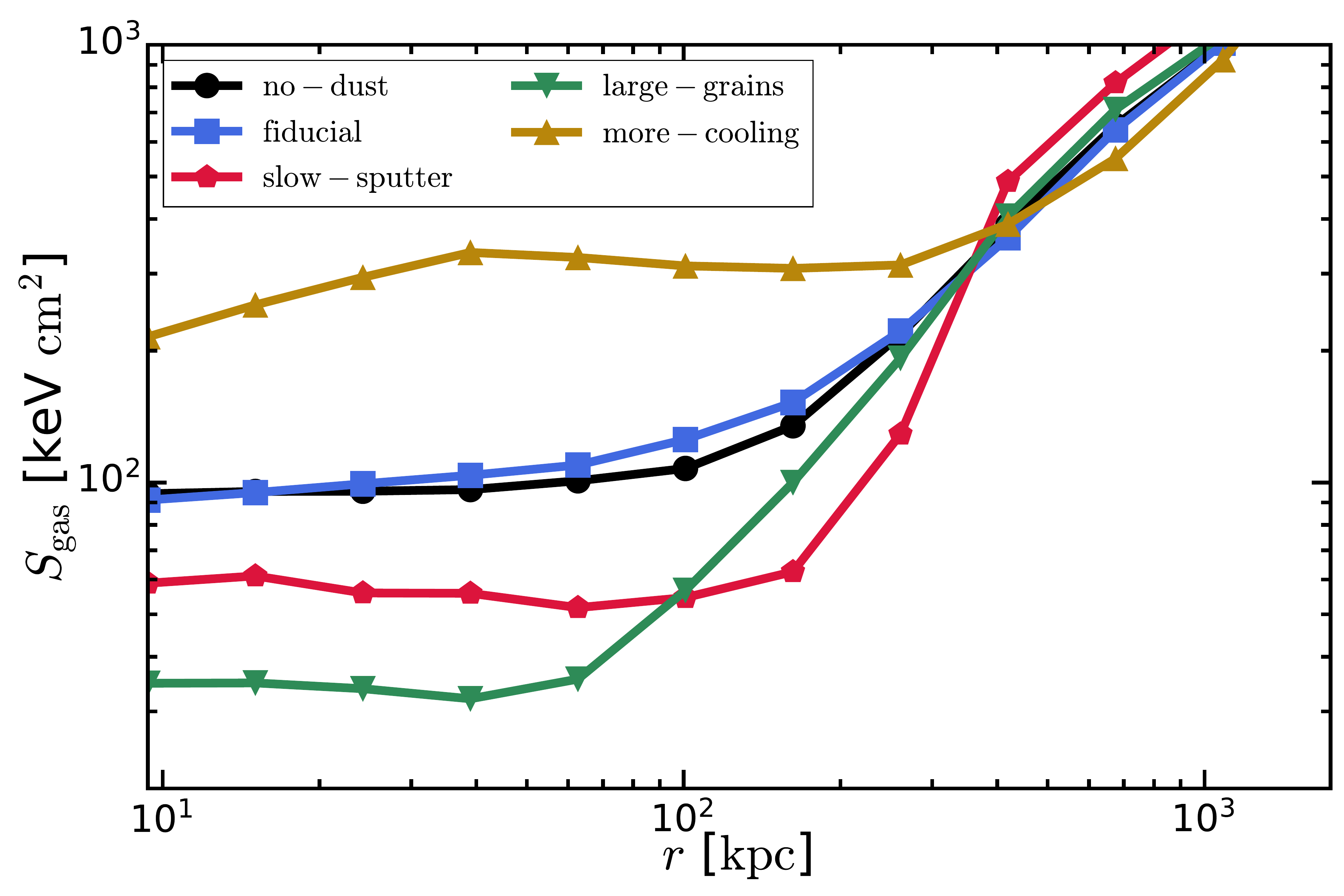}
\includegraphics[width=0.49\textwidth]{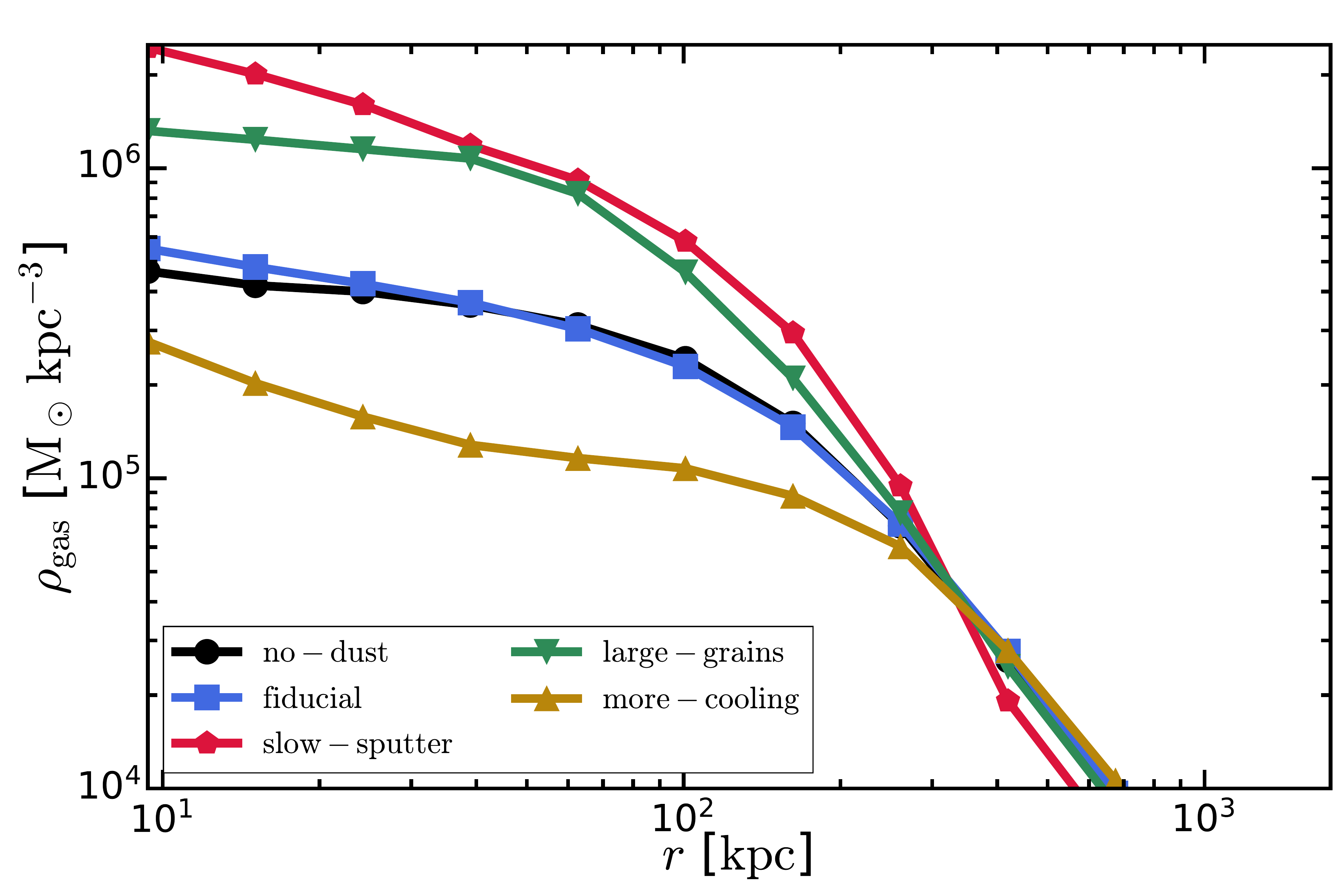}\\
\includegraphics[width=0.49\textwidth]{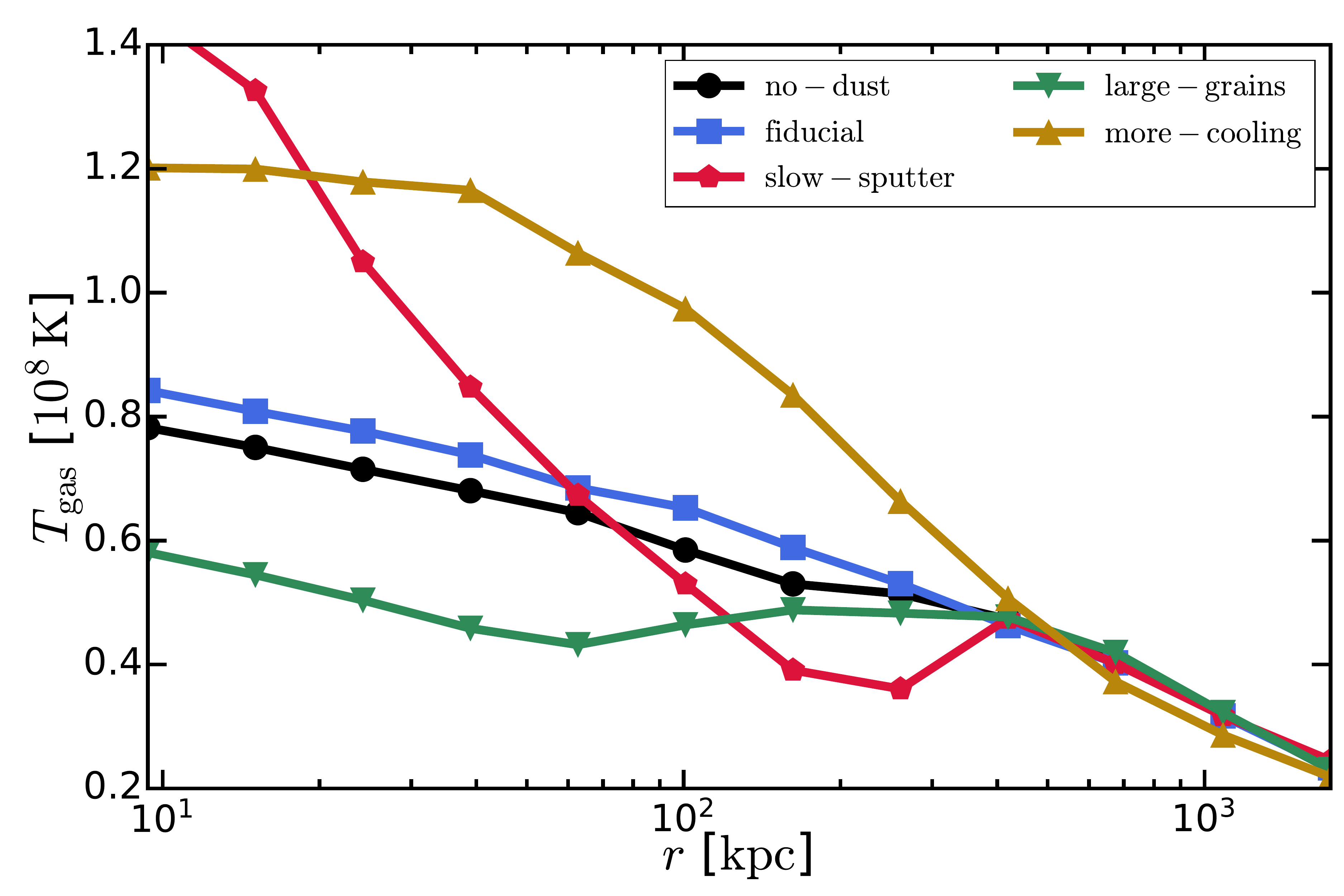}
\includegraphics[width=0.49\textwidth]{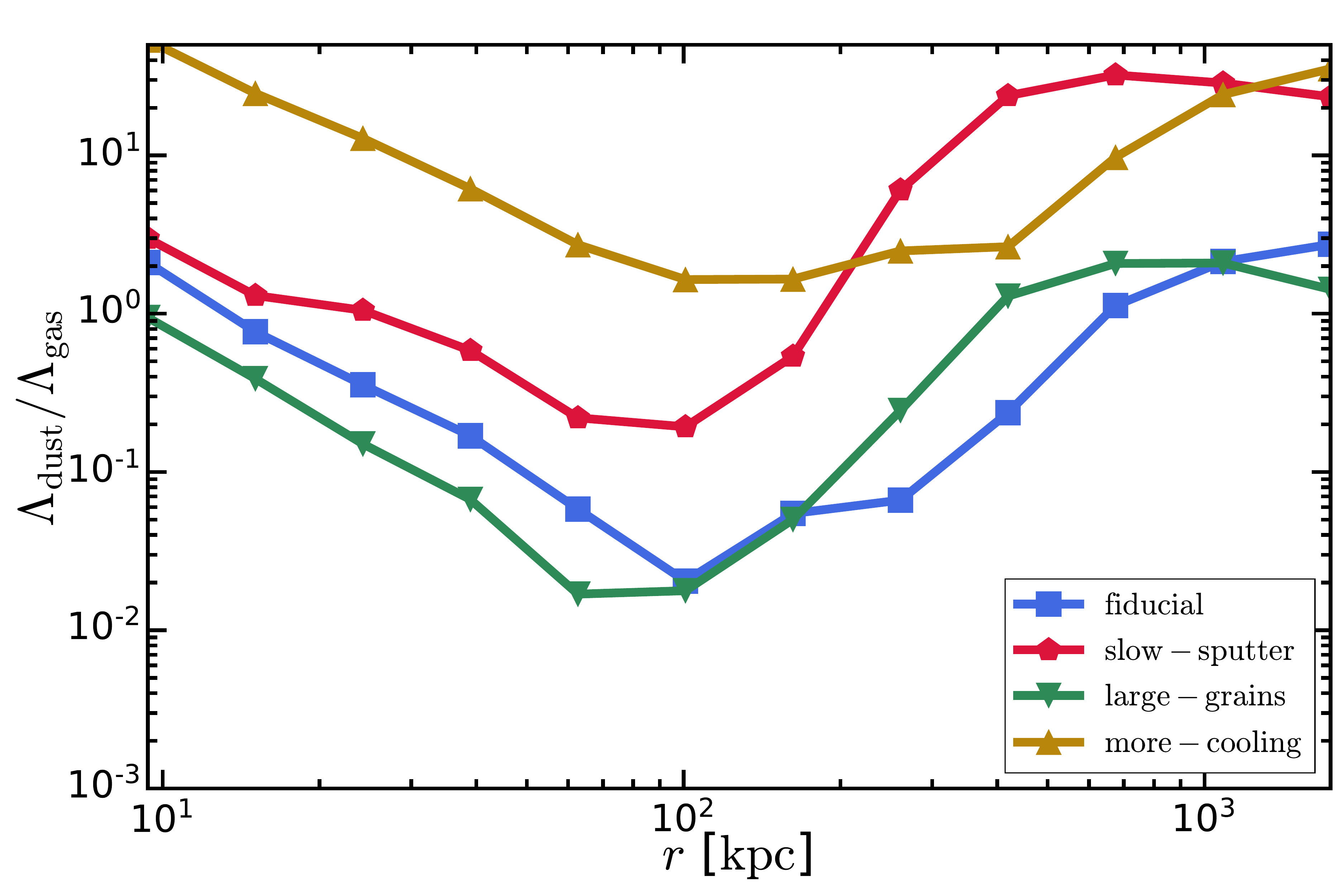}
\end{tabular}
\caption{{\bf Cluster gas radial profiles at $\mathbf{z=0}.$} {\it Upper left
panel:} Entropy profile. {\it Upper right panel:} Density profile.  {\it Lower
left panel:} Temperature profile. {\it Lower right panel:} Profile of the
cooling rate ratio of dust cooling over gas cooling. Gas cooling includes here
both primordial and metal line cooling. The inclusion of dust changes, in
general, the various thermodynamic profiles. The changes are driven mainly by
three coupled effects: the reduction of gas phase metal-line cooling due to gas
to dust condensation; the addition of dust IR cooling as additional coolant;
changes in the AGN feedback energy due to differences in the overall gas
accretion towards the center of the cluster. The combined outcome of these
effects is difficult to predict since the various effects operate in different
directions. Nevertheless, we find that this combined effect leads to noticeable
differences in the thermodynamic profiles when comparing the dust simulations
with the simulation without any dust physics.}
\label{fig:gas_profiles}
\end{figure*}

Based on our {\small\sc fiducial} dust model, we can infer from the thermal sputtering profiles that dust in the ICM should have
a typical sputtering timescale of about $\sim 10\Myr$ in the inner halo within
$\sim 100-200\kpc$. In the outer part, these timescales get much longer so that
at $1\Mpc$ cluster-centric distance, the sputtering timescale is already larger
than $1\Gyr$. This inferred thermal sputtering timescale is in rough
agreement with other estimates for this timescale. For example, \cite{Draine1979b} originally
estimated
\begin{equation}
\tau_{\rm sputter} = 2 \times 10^4\,{\rm yr} \, \left(\frac{{\rm cm}^{-3}}{n_{\rm H}} \right) \left(\frac{a}{0.01\mum} \right ) \nonumber
\end{equation}
for the thermal sputtering rate of dust grains in hot gases. For our fiducial
grain size of $0.1\mum$ and a rough hydrogen number density of $n_{\rm H} \sim
10^{-3}\cm^{-3}$ we find $\tau_{\rm sputter} \sim 100\Myr$, which is a typical
sputtering rate of our {\rm fiducial} model based on
Figure~\ref{fig:tausputter}.  As described above our thermal sputtering
timescale is also consistent with the dust thermal sputtering timescales of the
best-fit model in~\cite{Gjergo2018}. We stress however again that these sputtering
timescales have been increased compared to the fiducial values of \cite{McKinnon2017}.

Knowing how the sputtering rates change as a function of radius allows us 
to qualitatively understand the shapes of the dust-to-gas and dust-to-metal
ratio profiles discussed above. The fact that the amount of dust increases
towards the outer parts of the cluster is due to the very long thermal
sputtering timescales in that part of the ICM, where dust can then survive longer.  Looking only at the sputtering timescales one would then naively
also expect that the amount of dust should be very low within $\sim
100-200\kpc$.  In fact, the minima of the ratio profiles occur at around
these radii, but the dust-to-gas and dust-to-metal ratios then increase
again towards the cluster center. This is caused by stronger dust production
and dust growth towards the center of the cluster, where the central galaxy
provides conditions such that dust can also grow quite efficiently.
Therefore, despite the fact that the sputtering timescales in the center are as
short as about $\sim 10\Myr$, we find that the dust-to-gas and
dust-to-metal ratios increase towards the center due to an increased dust
production and growth in the inner regions. Therefore, the combination of
dust production and growth together with the shape of the thermal sputtering
timescale profiles explain the radial dependence of the dust-to-gas and
dust-to-metal profiles.

We conclude from this Section that the distribution of dust within the cluster is
consistent with existing observational data. Our {\small\sc fiducial} model predicts
a typical thermal sputtering timescale of about $\sim 10\Myr$ for the inner parts of
the ICM.

\section{Impact of dust on thermodynamic cluster profiles}
\label{sec:Section5}

So far, we have quantified the abundance of dust and its distribution in the
cluster. Our models predict a small amount of dust in agreement with current
observational constraints and estimates. Specifically, our {\small\sc fiducial}
model predicts an overall dust-to-gas ratio of about $\sim 2\times 10^{-5}$
averaged over the full cluster.  This small amount of dust in the ICM can
potentially also affect the thermodynamic state of the cluster gas due to dust
IR cooling and the reduction of gas phase metals due to metal to dust
condensation, which leads to a reduction of metal-line cooling.  Various works
have in the past studied the potential impact of dust on the
thermodynamic structure of galaxy
clusters~\citep[e.g.,][]{Montier2004,Pointecouteau2009,DaSilva2009,Melin2018}.
For example, \cite{Montier2004} predicted that dust IR cooling is important in the
ICM for gas temperatures ${T_\text{gas} = 10^6 - 10^8\K}$, and if $D > 2\times 10^{-5}$.
\cite{DaSilva2009} found a $25\%$ normalisation change for the $L_\text{X}-M$ relation
and $10\%$ change for the $Y-M$ and $S-M$ cluster scaling relations in the
presence of dust. Similarly, \cite{Pointecouteau2009} found changes in the
$L_\text{X}-M$ relation by as much as $10\%$ for clusters with temperatures
around $1\,{\rm keV}$ for models that include dust cooling.  However, these
results are based on rather crude and limited dust models. We will therefore inspect the impact of dust on the thermodynamic profiles of the
ICM for our dust model in more detail in this Section.

There are mainly two different mechanisms through which dust can affect the
thermodynamics of the cluster gas.  First, the condensation of gas phase metals
into dust grains reduces the strength of metal line cooling due to the reduced
metal budget in the gas.  Second, the presence of dust in the hot ICM causes IR
cooling of the gas. Both of these effects act in opposite directions; i.e. it is
a priori not clear how the addition of dust to the ICM affects the overall
cooling.  Another complication is the backreaction of these cooling changes on
the feedback mechanisms in the cluster. Most importantly, the accretion rate
onto the central supermassive black hole (SMBH) is sensitive to the gas
cooling, which regulates how much gas can flow towards the cluster center. We
therefore expect that the change of gas cooling due to dust will also impact
the accretion rate onto the supermassive black hole and consequently the active
galactic nuclei (AGN) feedback.  All these effects are coupled in a non-linear
way, making it difficult to predict the final impact of dust on the
thermodynamic structure on the cluster.  Given this difficult interplay it is
crucial to study this problem through numerical simulations.  Before discussing
our results, we stress that our work represents only an initial
exploration of these effects since our sample contains only one cluster. The
impact of dust depends most likely on halo mass as well as details of the
formation history. One has to keep these limitations in mind for the following.

In Figure~\ref{fig:gas_profiles} we present the basic thermodynamic profiles of
the ICM for the different dust models: entropy profiles (upper left), gas
density profiles (upper right), gas temperature profiles (lower left). We also present in the lower right panel the ratio of dust IR cooling
and gas cooling. Gas cooling here includes both primordial and metal line
cooling. For the entropy, density and temperature profiles we only consider
non-star-forming gas in the ICM.  All thermodynamic profiles demonstrate that
dust physics indeed has an impact on the thermodynamic state of the ICM gas.
However, there seems to be no simple one-to-one mapping from included dust
physics to the final outcome of the thermodynamic profiles. This has to do with
the non-linear coupling between the different effects caused through the
inclusion of dust as described above. Two of our dust models, {\small\sc large-grains} and {\small\sc slow-sputter}, lead to an
overall reduction of the central entropy compared to the {\small\sc no-dust} model.
The {\small\sc more-cooling} model, on the other hand, leads to a significantly higher central entropy
compared to the {\small\sc no-dust} case.  This is caused by an increased
central temperature and reduced central density as can be seen in the other
panels. We have also inspected the injected AGN energy, which is higher for
this model; i.e. the central gas is more heated compared to the other models
due to the increased gas accretion rate caused by the large amount of dust
cooling occurring for that model. As mentioned above, the competition between
altered central AGN heating, dust-induced reduced gas phase metal line cooling,
and dust IR cooling leads to the large variety of thermodynamic profiles once
dust physics is included in the simulation. Despite these complications, we
conclude that in any case dust can alter the thermodynamic profiles of the cluster. Interestingly,
the different dust driven effects lead to a nearly unchanged entropy profile for the {\small\sc fiducial} model.  We note
that the stellar mass of the central galaxy is also affected by the presence of dust. The {\small\sc slow-sputter} dust model causes the
largest impact on the stellar mass with an overall increase of stellar mass by a factor of $\sim 3$ compared to the {\small\sc no-dust} model. 

\begin{figure*}
\centering
\includegraphics[width=1\textwidth]{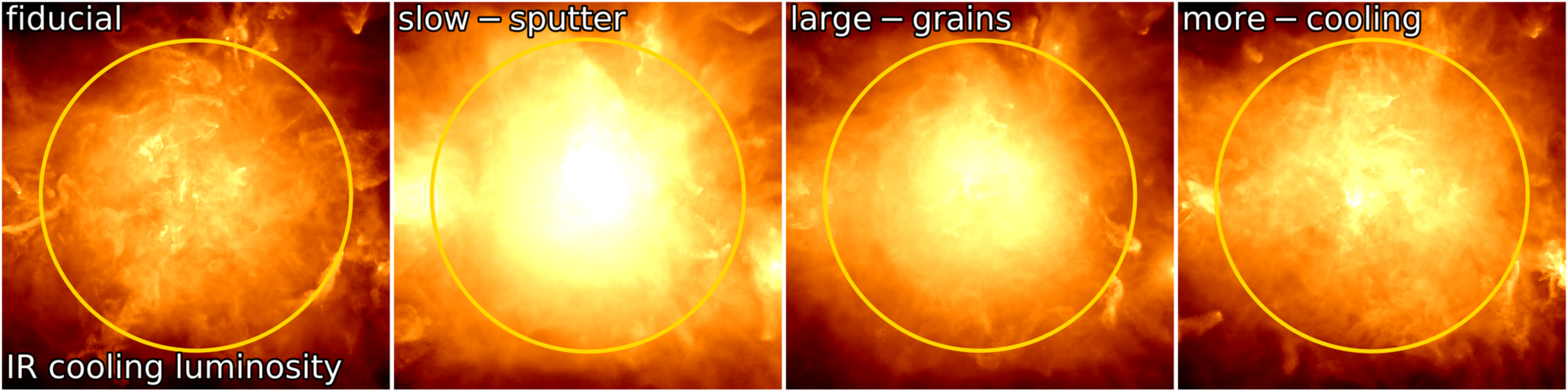}
\includegraphics[width=0.235\textwidth]{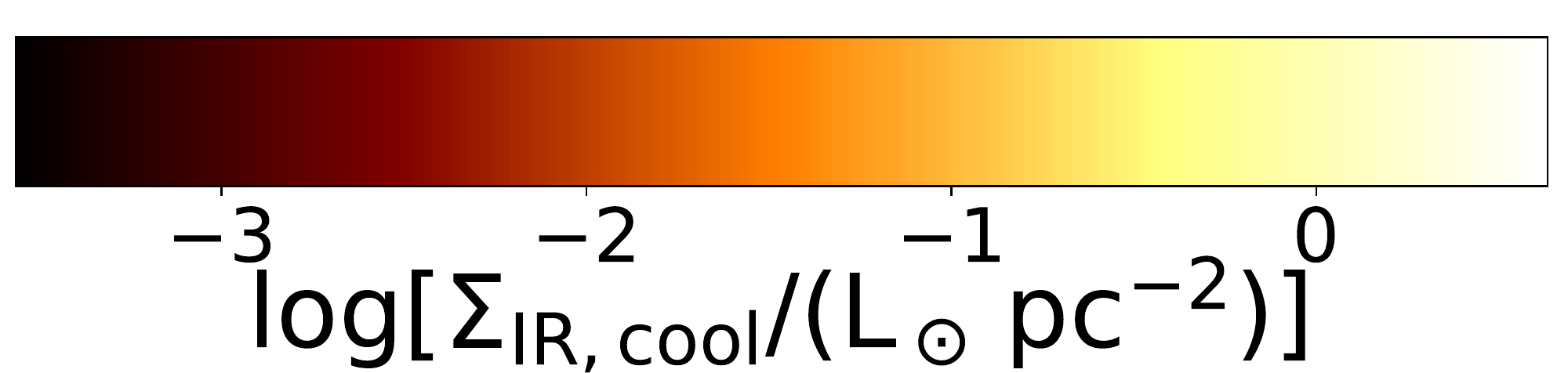}
\caption{{\bf Dust IR cooling luminosity maps at $\mathbf{z=0}$.} The cooling
of high temperature gas in the presence of dust emits IR radiation. The circles show $r_{\rm 200,
crit}$. The IR dust
cooling radiation is strongest for the model with low thermal sputtering rates,
{\small\sc slow-sputter} due to the large amounts of dust in the ICM. The IR
emission is lower for the {\small\sc fiducial} model, due to the lack of dust
in the ICM. The emission of the remaining two dust models fall in between those
two.}
\label{fig:dust_maps_2}
\end{figure*}

The effect of reduced metal line cooling due to dust-to-metal condensation can
be inferred indirectly through the dust-to-metal profiles presented
in~Figure~\ref{fig:dust_to_gas}.  However, some of this reduction in metal line
cooling is compensated by the additional dust IR cooling that we
discuss next. The strength of this dust IR cooling is presented in the lower
right panel of Figure~\ref{fig:gas_profiles}, where we present the ratio of dust
IR cooling over the sum of primordial and metal line cooling.  The various dust
models lead to a large variation in the amount and distribution of dust in the
ICM as demonstrated above. It is therefore not surprising to see strong variations in the contribution of dust to the overall gas cooling in the
cluster. For example, the contribution from dust IR cooling can vary by more
than two orders of magnitudes for gas at around $\sim 100\kpc$ cluster-centric
distance.  The question then arises of how much dust cooling we can realistically
expect in a cluster, and how large is its impact on the thermodynamic profile
of the cluster compared to the case where we do neglect all dust physics.  We
have demonstrated above that essentially all dust models, except the {\small\sc slow-sputter} model, agree reasonably well with
most observational constraints. We can therefore expect that these models roughly bracket the potential impact of dust
on the thermodynamic cluster profiles. 

The cooling ratio panel demonstrates that dust cooling is most relevant
in the innermost and outermost parts of the cluster. In fact, in the outer part
the cooling due to dust IR emission can overcome the metal line contribution for
the {\small\sc fiducial} model. For the model with
even lower and most likely unrealistic sputtering rates, {\small\sc slow-sputter}, we find that the dust
cooling rate is at most radii significantly larger than the metal line cooling
rate. Especially at larger radii we find that the dust cooling is more than a
factor of $20$ larger then metal-line cooling of the gas. The model with
larger grains, {\small\sc large-grains}, also leads to a slightly increased cooling rate at most radii. We note
that in this case multiple physical effects are altered once the grain size is changed.
First, dust can survive much longer in the ICM due to the increase sputtering timescale, i.e. its abundance and consequently the
dust-to-gas ratio increase substantially. Second, the growth also slows down, but
the net effect of the reduced growth and reduced sputter is still a substantially
increased dust-to-gas ratio in the ICM as demonstrated in Figure~\ref{fig:dust_to_gas}. Third,
the cooling rate itself is also sensitive to the grain size. And last, for larger grains the gas phase metal abundance is also strongly reduced in the ICM
due to the longer survival of larger grains in this hot environment. Therefore, the gas metal line cooling is also
reduced. 
Obviously, the model with five times larger dust cooling rates also has a larger
ratio between dust cooling and metal-line cooling compared to the {\small\sc
slow-sputter} model. 

The dust cooling process works through the emission of IR radiation from the
dust to radiate away the energy of the gas. We can therefore use cluster IR measurements to quantify 
whether the bolometric IR cooling luminosity predicted by our model is
consistent with those measurements or exceeds it. Since
we are only interested in the bolometric dust IR luminosity, we do not need any
information about the IR spectral shape, which would require a model to track
the dust temperature, which is not included in our simulations.  For the
bolometric dust IR luminosity we can simply sum up the dust cooling
rate $\Lambda_{\rm dust}(T_\text{gas})$ within the cluster and assume that this cooling
rate is equivalent to the bolometric dust IR emission. We present maps of this
dust cooling IR radiation for our different dust models in
Figure~\ref{fig:dust_maps_2}. We note that these bolometric IR maps only account
for the cooling emission due to dust, and do not take into account other
sources of IR emission. As expected, the emission is strongest for the
{\small\sc slow-sputter} model, which contains the largest amount of dust in
the ICM. The {\small\sc fiducial} model emits much less IR radiation. In those
maps we can also see the emission of stripped dust more clearly. Larger grains
also lead to slightly more cooling emission compared to the {\small\sc
slow-sputter} case.  Boosting the dust cooling efficiency, as in the {\small\sc 
more-cool} model, obviously also increases the amount of IR emission.

\begin{figure}
\centering
\includegraphics[width=0.495\textwidth]{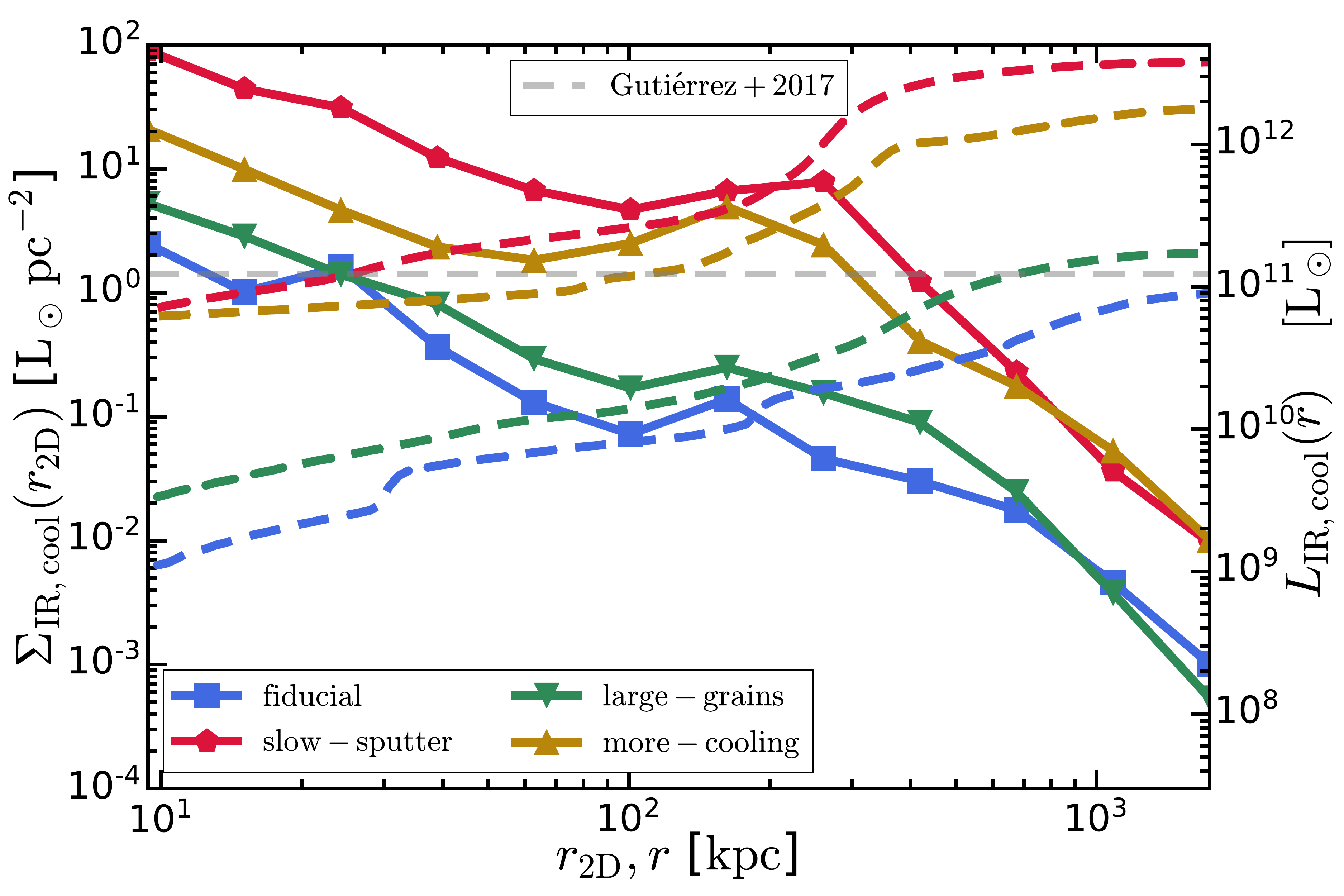}
\caption{{\bf Surface IR luminosity profiles of the dust cooling radiation and
cumulative IR cooling radiation luminosity at $\mathbf{z=0}$.} This radiation only includes
IR cooling emission from dust; i.e. all other IR sources are not included in the
model calculations. We also include observational data for the total cluster IR luminosity
taken from~\protect\cite{Guti2017}. }
\label{fig:luminosity_profile}
\end{figure}

The dust IR emission is quantified in more detail in
Figure~\ref{fig:luminosity_profile}, where we present both IR surface brightness
profiles (solid lines), and cumulative dust luminosity profiles (dashed lines)
for the different dust models. We only include non-star-forming gas here. For
the total luminosity we also show some observational results
from~\cite{Guti2017}, who measured the ICM IR emission using a sample of
$327$ clusters of galaxies subtracting the contribution of identified sources
from the whole emission of the clusters. Their massive low redshift sample has
an average cluster mass of $1.59\times 10^{14}\msun$ and an average redshift of
$z=0.173$ with a range of $z=0.06-0.24$.  For this high mass sample, they find
an IR luminosity of $4.7 \times 10^{44}\,{\rm erg}\,{\rm s}^{-1}$. We note that their derivation of the bolometric IR luminosity
depends on some assumptions about, for example, the dust temperature in the ICM,
which introduces some systematic uncertainties in the derived value. Comparing
their derived luminosity with our dust model results, we find that 
the {\small\sc slow-sputter} and {\small\sc more-cooling} produce too much IR radiation. However, we note that the average mass of the
observational cluster sample is slightly lower than the mass of our simulated cluster. It
is therefore possible that even the {\small\sc more-cooling} model could
still be consistent. We therefore conclude that the IR dust emission of all our
dust models, except {\small\sc slow-sputter}, are consistent with these observational findings.
We note that radiative dust heating due to the UV and Cosmic Microwave
Background can be neglected here, given the low radiation field in the intracluster medium~\citep[][]{Montier2004}.

We can also compare the total IR luminosities due to dust cooling to theoretical estimates.
For example, \cite{Montier2004} estimated analytically that the total IR dust cooling emission
should be a function of the dust-to-gas ratio $D$ and the electron density $n_{\rm e}$ in the ICM given by:
\begin{equation}
L_{\rm IR, cool} \! = \! 3.7 \! \times \! 10^{45}{\rm erg}\,{\rm s}^{-1} \! \left(\frac{D}{10^{-5}}\right) \! \left(\frac{n_{\rm e}}{10^{-3}\cm^{-3}}\right) \! \times \! G_{\rm eom}. \nonumber
\end{equation}
Here $G_{\rm eom}$ is a geometrical factor. If we assume that ${\rho_{\rm gas}
\sim 10^4\msun\kpc^{-3}}$ and ${D\sim 10^{-5}}$ as typical values, we find roughly ${L_{\rm IR,cool}
\sim 10^{11}\,{\rm L}_\odot \times G_{\rm eom}}$, which is in reasonable agreement with the results of
our model.

We conclude from this Section that the small amounts of dust in the ICM alter
the thermodynamic profiles of the ICM due to changes in the overall cooling rate. The dust IR
cooling emission is consistent with current cluster IR measurements and also with previous
theoretical estimates. 

\section{Conclusions}
\label{sec:Section6}

We have performed cosmological hydrodynamical simulations to study the dust
content of a galaxy cluster, ${M_\text{200,crit}=6 \times 10^{14}\msun}$, using
a novel self-consistent dust model including dust production, growth,
supernova-shock-driven destruction, ion-collision-driven thermal sputtering,
and high temperature dust cooling through far infrared re-radiation of
deposited electron energies. This dust model is coupled to a galaxy formation
model implemented in the moving-mesh code {\sc Arepo}. Our main results are:

\begin{itemize}[leftmargin=*]
\item Our {\small\sc fiducial} dust model, which employs a reduced thermal sputtering rate
compared to our \cite{McKinnon2017} sputtering parametrisation, reproduces the observed dust
abundances in clusters at low redshifts. The average dust-to-total-mass and
dust-to-gas ratios decrease as a function of time and reach their lowest values
at redshift $z=0$. The dust mass itself, on the other hand, increases at high $z$,
has a broad maximum at around $z\sim 1.5 -  2$, and then declines again
towards lower redshifts. Our {\small\sc fiducial} model predicts at the present
day a total dust mass fraction of $\sim 3 \times 10^{-6}$ within $r_{\rm
200,crit}$. Within the same radius, the model also predicts at $z=0$ a
dust-to-gas ratio of $\sim 2\times 10^{-5}$. These values correspond to an
absolute dust mass of about $2 \times 10^{9}\msun$. All these values are
consistent with current observational data and constraints. The peak dust mass
at higher redshifts around $z \sim 1.5 - 2$ is close to $10^{11}\msun$.

\item For our {\small\sc fiducial} model, thermal sputtering timescales for the
inner $\sim 100\kpc$ of the ICM are approximately $10\Myr$. The sputtering
timescales then rise towards the outskirts of the ICM such that at $1\Mpc$
cluster-centric distance, the timescales are of the order of $\sim 1000\Myr$.
These sputtering timescales are consistent with those recently found by
\cite{Gjergo2018}. 

\item Our {\small\sc fiducial} dust model predicts a dust-to-metal ratio of
${\sim 10^{-2}}$ at lower redshifts, with ratios increasing towards higher
redshifts. All our models predict a dust-to-metal ratio close to unity at
${z\sim 4}$. A model with a ten times longer thermal sputtering timescale,
{\small\sc slow-sputter}, predicts at $z=0$ a ten times larger dust-to-metal
ratio of about $\sim 10^{-1}$.

\item Most of our dust models predict very flat
dust surface density profiles consistent with observational findings. Our
{\small\sc fiducial} model yields a dust surface density slightly below ${\sim
10^{-3}\msun\pc^{-2}}$.  The model with a ten times lower thermal sputtering
rate, {\small\sc slow-sputter}, results in an about $\sim 10$ higher dust
surface density.

\item The dust-to-gas and dust-to-metal radial profiles show a characteristic
shape with minima occurring at around $\sim 100-200\kpc$
cluster-centric distance and rising ratios towards the center and outskirts of
the cluster. The increase towards the outskirts is due to the smaller thermal
sputtering rates caused by lower gas temperatures
and densities in the outer parts of the ICM. Towards the inner parts, the thermal sputtering rates are higher,
but dust production and growth in these regions is also larger. The combined
effect of increased production and growth rates towards the center and the
reduced sputtering towards the outer parts lead to the characteristic profile
shape of the dust-to-gas and dust-to-metal ratio profiles. Our {\small\sc
fiducial} model predicts a central dust-to-gas ratio of $\sim 10^{-5}$ and a
minimum ratio of $\sim 10^{-7}$ at intermediate radii. The ratio increases 
again towards larger radii and reaches a value of about $\sim 10^{-5}$ at
$\sim 1\Mpc$ cluster-centric distance. 

\item Dust has an impact on the thermodynamic profiles of the cluster 
gas. This is caused by the rather complicated interplay of three effects.
First, the formation of dust occurs at the expense of gas-phase metals such
that metal-line cooling rates are reduced. Second, dust itself is an efficient
coolant in hot plasma due to the emission of IR radiation. These two effects
work in opposite directions on the total cooling rates. Third, the overall change in the
cooling rate also implies a change in the
AGN feedback activity due to the change in the accretion rates onto the SMBH. These three
effects combine and affect the thermodynamic state of the ICM. 
We stress, however, that these results are based on a single halo.
Given the non-linear coupling between the three effects discussed above, it is
likely that the impact can fluctuate quite strongly from halo to halo.

\item Dust cooling operates through the emission of IR radiation. The predicted
IR bolometric luminosities by our {\small\sc fiducial} model are consistent with current constraints from cluster
IR measurements. Our {\small\sc fiducial} model produces a bolometric IR
luminosity of about $10^{10}\,{\rm L}_\odot$ within ${\sim 1\Mpc}$. A model
with reduced thermal sputtering rates, {\small\sc slow-sputter}, results in a
nearly $\sim 100$ times higher IR luminosity.
\end{itemize}

We conclude that, according to our simulations, small amounts of dust should be
present in ICM despite the hostile environment for dust survival. However, the
exact amount of dust is very sensitive to the details of the thermal sputtering
model. Our {\small\sc fiducial} model is consistent with current observational
dust constraints. However, this agreement is achieved mostly
due to a reduced thermal sputtering rate. Further studies are required to fully
quantify this phenomena in more detail exploring a wider halo mass range. We also expect that
dust formation has a significant effect in lower mass halos by
altering the gas cooling rates and changing gas-phase metal abundances in the
circumgalactic medium due to dust-to-metal condensation.

\section*{Acknowledgements}
We thank the anonymous referee for very useful feedback. We thank Volker Springel for giving us access to {\sc Arepo}.
The simulations were performed on the joint MIT-Harvard computing cluster
supported by MKI and FAS. MV acknowledges support through an MIT RSC award, a
Kavli Research Investment Fund, NASA ATP grant NNX17AG29G, and NSF grants
AST-1814053 and AST-1814259.  RM acknowledges support from the DOE CSGF under
grant number DE-FG02-97ER25308. RK acknowledges support from NASA through Einstein Postdoctoral 
Fellowship grant number PF7-180163 awarded by the Chandra X-ray Center, 
which is operated by the Smithsonian Astrophysical Observatory for NASA 
under contract NAS8-03060.

\label{lastpage}

\end{document}